\documentclass[final,3p,times]{elsarticle}

\usepackage{amssymb}
\usepackage{amsthm}
\usepackage{makeidx}  
\usepackage{algorithm}
\usepackage[noend]{algorithmic}
\usepackage{qtree}
\usepackage[table]{xcolor}
\usepackage{graphicx}
\usepackage{graphics}
\usepackage{amsmath}
\usepackage{amsfonts}  
\usepackage{parskip}
\usepackage{verbatim}
\usepackage{amsmath}
\usepackage{amsfonts}
\usepackage{multicol}
\usepackage{latexsym}
\usepackage{booktabs}
\usepackage{color}
\usepackage{qtree}
\usepackage{tikz}
\usepackage{tikz-er2}

\def\punto{\hspace*{\fill}\Box}

\usetikzlibrary{arrows}
\usetikzlibrary{positioning,chains,fit,shapes,calc}
\usetikzlibrary{shadows}

\tikzstyle{every entity} = [top color=white, bottom color=blue!30, draw=blue!50!black!100, drop shadow]
\tikzstyle{every weak entity} = [drop shadow={shadow xshift=.7ex, shadow yshift=-.7ex}]
\tikzstyle{every attribute} = [top color=white, bottom color=yellow!20, draw=yellow, node distance=7em, drop shadow]
\tikzstyle{every relationship} = [top color=white, bottom color=red!20, draw=red!50!black!100, drop shadow]
\tikzstyle{every isa} = [top color=white, bottom color=green!20, draw=green!50!black!100, drop shadow]

\usepackage{listings}

\usepackage{color}
\definecolor{gray}{rgb}{0.4,0.4,0.4}
\definecolor{darkblue}{rgb}{0.0,0.0,0.6}
\definecolor{cyan}{rgb}{0.0,0.6,0.6}

\lstdefinelanguage{XML}
{
  basicstyle=\ttfamily\color{darkblue}\bfseries,
  morestring=[b]",
  morestring=[s]{>}{<},
  morecomment=[s]{<?}{?>},
  stringstyle=\color{black},
  identifierstyle=\color{darkblue},
  keywordstyle=\color{cyan},
  morekeywords={xmlns,version,type}
}

\def\R{{\mathbb R}}

\biboptions{sort&compress}

\journal{Knowledge Based Systems}
\begin{document}
\begin{frontmatter}

\newtheorem{definition}{Definition}
\newtheorem{theorem}{Theorem}[section]
\newtheorem{lemma}[theorem]{Lemma}
\newtheorem{proposition}[theorem]{Proposition}
\newtheorem{corollary}[theorem]{Corollary}
\newtheorem{example}{Example}[section]

\hyphenation{}

\title{XML Matchers: approaches and challenges}

\author[1]{Santa Agreste}
\address[1]{Department of Mathematics and Informatics, University of Messina, I-98166 Messina, Italy}
\ead{sagreste@unime.it}
\author[2]{Pasquale De Meo}
\address[2]{Department of Ancient and Modern Civilizations, University of Messina, Italy.}
\ead{pdemeo@unime.it}
\cortext[cor1]{Corresponding author}
\author[3]{Emilio Ferrara}
\address[3]{School of Informatics and Computing, Indiana University Bloomington, USA}
\ead{ferrarae@indiana.edu}
\author[4]{Domenico Ursino}
\ead{ursino@unirc.it}
\address[4]{Dept. of Information, Infrastructure and Sustainable Energy Engineering, University Mediterranea of Reggio Calabria, I-89122 Reggio Calabria, Italy}

\address{}

\begin{abstract}
Schema Matching, i.e. the process of discovering semantic correspondences between concepts adop\-ted in different data
source schemas, has been a key topic in Database and Artificial Intelligence research areas for many years. In the
past, it was largely investigated especially for classical database models (e.g., E/R schemas, relational databases,
etc.). However, in the latest years, the widespread adoption of XML in the most disparate application
fields pushed a growing number of researchers to design XML-specific Schema Matching approaches, called XML Matchers,
aiming at finding semantic matchings between concepts defined in DTDs and XSDs. XML Matchers do not just take
well-known techniques originally designed for other data models and apply them on DTDs/XSDs, but they exploit specific
XML features (e.g., the hierarchical structure of a DTD/XSD) to improve the performance of the Schema Matching process.
The design of XML Matchers is currently a well-established research area. The main goal of this paper is to provide a
detailed description and classification of XML Matchers. We first describe to what extent the specificities of
DTDs/XSDs impact on the Schema Matching task. Then we introduce a template, called {\em XML Matcher
Template}, that describes the main components of an XML Matcher, their role and behavior. We illustrate how each of
these components has been implemented in some popular XML Matchers. We consider our XML Matcher Template as the
baseline for objectively comparing approaches that, at first glance, might appear as unrelated. The introduction of
this template can be useful in the design of future XML Matchers. Finally, we analyze commercial tools implementing XML
Matchers and introduce two challenging issues strictly related to this topic, namely XML source clustering and
uncertainty management in XML Matchers.
\end{abstract}


\end{frontmatter}

\section{Introduction}
\label{sec:intro}

The eXtensible Markup Language (XML) has emerged as a de-facto standard for the representation and the exchange of data
in a wide range of scenarios \cite{Thompson*01,Ursino-IS4,Lee*02}.

As an example, XML has been widely adopted in many scientific domains, like biology \cite{Hucka*03}, chemistry
\cite{Murray97} and geography/geology \cite{Cox01}, to name a few. In the context of e-commerce, many Web sites use XML
as a tool to encode their catalogue of products, as well as to represent business documents, like invoices or orders.

To make data exchange easier, organizations like the World Wide Web Consortium (W3C) are increasingly committed to the
definition of advanced languages to describe the structure and content of an XML data source. One of the first
languages was {\em XML DTD} (Document Type Definition) \cite{XML-DTD98}. Later, W3C developed a more complex language
called {\em XSD} (XML Schema Definition), also known as WXS (W3C Schema Definition), in order to overcome some
limitations of DTD \cite{XSD12}\footnote{For instance, the fact that DTD does not support namespaces and that data type
management provided by it is weak and can be applied only to attributes.}.

These languages are used to build a {\em schema}, i.e. a collection of metadata called {\em schema components} or {\em
schema elements}. A schema specifies a set of rules an XML document must obey in order to be considered {\em valid}
according to the schema itself. The availability of a schema significantly simplifies data exchange procedures
\cite{Ursino-IS4}. For instance, there exist simple software programs that can check if a given document satisfies the
constraints imposed by a given schema, and, in the positive case, allow for seamless data exchange among the interested
parties.

Despite the presence of powerful languages like DTD/XSD, the achievement of the full interoperability among
applications based on XML data is often illusory. DTDs/XSDs, in fact, provide {\em self-describing} capabilities, i.e.
they allow designers to define names for elements and attributes. However, the widespread assumption that these names
denote some intrinsic semantics can be false; therefore, it is not sufficient to look at element/attribute names to
catch the content of an XML document \cite{Ehrig06,WiGl08}.

An indefeasible precondition to achieve interoperability consists of detecting and identifying if two or more schema
elements convey the same semantics, despite, for instance, they have different names.

The task of finding pairs (or groups) of elements sharing the same semantics has a long tradition in Computer Science:
In Database and Artificial Intelligence literature such a problem is known as {\em Schema Matching} \cite{RaBe01},
whereas in the Semantic Web community it is known as {\em Ontology Alignment} \cite{Ehrig06} or {\em Ontology Matching}
\cite{EuSh07}. The vast majority of these approaches was reviewed in books or surveys
\cite{RaBe01,BeMaRa11,BeBoRa11,Ehrig06,EuSh07}. However, most of them do not make any specific assumption on the data
model and format used to represent Schemas/Ontologies. Most of them have been originally designed to work on some data
models (e.g., E/R diagrams) but, subsequently, have been applied and tested on DTD/XSD.

However, in the latest decade, a growing number of researchers focused on the specificities of DTD/XSD and considered
them in the design of new, ad-hoc, Schema Matching approaches. We call {\em XML Matchers} these approaches.

DTDs/XSDs offer advanced capabilities and features that are not present in other data models. The usage of these
features is far from trivial, but it can have a positive impact in the design of an XML Matcher. By now, Schema
Matching of XML sources is a mature and well-established research area in which several and authoritative contributions
have been provided.

The goal of this survey is to offer a comprehensive coverage of XML-specific Schema Matching, regarded as a narrow
sub-area of Schema Matching.

The main contributions of this paper are as follows:

\begin{enumerate}

\item We describe to what extent the specificities of DTD/XSD impact on the Schema Matching process. We show that
    the hierarchical features of DTD/XSD open up new research problems that do not emerge, for instance, in the
    matching of E/R diagrams or relational schemas.

\item We provide a template, called {\em XML Matcher Template}, describing the main components of an
    XML Matcher, as well as their role and behavior. We discuss how each of these components has been implemented
    in some popular XML Matchers. This helps us to better describe how XML Matchers work in practice. Our template
    acts as a tool for highlighting and understanding commonalities among XML Matchers that, at a first glance,
    could appear as totally unrelated. It can act as a baseline for future work allowing research on this topic to
    make progress at a faster pace.

\item  We discuss some commercial prototypes designed to find matchings between DTDs/XSDs and we use our XML
    Matcher template to classify them.

\item We discuss two challenges strictly related to XML-specific Schema Matching, namely XML source clustering and
    uncertainty management in XML Matchers.

\end{enumerate}

This paper is structured as follows: In Section \ref{sec:related-work} we review related surveys on Schema
Matching/Ontology Matching with the aim of showing the main novelties brought in by this work. In Section
\ref{sec:schema-matching-review} we summarize the basic notions and definitions about Schema Matching. Our XML Matcher
Template is presented in Section \ref{sec:matching-xml}, whereas Section \ref{sec:conceptual} provides a framework to
systematically classify XML-specific Schema Matching approaches. In Section \ref{sec:commercial} we provide an overview
of commercial XML Matchers. In Section \ref{sec:Challenges} we examine two important challenges related to XML Matcher
research, namely XML source clustering and uncertainty management in XML Matchers. In Section \ref{sec:discussion} we
discuss the main lessons learned from our analysis. Finally, in Section \ref{sec:conclusions}, we draw our conclusions.

\section{Related work}
\label{sec:related-work}

Schema Matching techniques have been studied in a large variety of application contexts, like data integration,
e-commerce, Data Warehousing, distributed query answering, to name a few \cite{BeBoRa11,RaBe01}.

Up to 2001, Schema Matching was considered as an issue functional to a specific application domain. Such a vision was
overturned by Rahm and Bernstein \cite{RaBe01}. They analyzed existing literature and recognized relevant similarities
among techniques which were originally designed to work in different application domains. As a consequence, they
suggested to consider Schema Matching as a new research problem which was interesting {\em per se}, independently of a
particular application domain. The classification criteria illustrated in \cite{RaBe01} were (and still are) warmly
welcomed by researchers working in the Schema Matching field, and they have been largely exploited to categorize
existing approaches.

An update of the work presented in \cite{RaBe01} is proposed in \cite{BeMaRa11}. In that paper the authors report the
main developments in Schema Matching algorithms in the decade 2001-11 and suggest a list of open research problems and
current research directions in the Schema Matching field.

A further, excellent survey on Schema Matching can be found in the book edited by Bellahsene et al. \cite{BeBoRa11}.
However, the topics discussed therein differ from the material covered in this survey; in fact, a relevant part of the
material in that work focuses on the usage of semantic matchings to perform schema evolution and schema merging; this
topic is not considered in this survey. In addition, some chapters are devoted to describe the metrics adopted for
experimentally assessing the performance of a Matcher as well as the strategies to tune a Matcher in such a way as to
optimize its efficiency or the quality of discovered matchings. Due to space limitations, we do not discuss these
problems in this survey, even if most of them are still valid in the context of XML-specific Schema Matching. This
survey introduces several novelties with respect to the book by Bellahsene et al. \cite{BeBoRa11}. First, it focuses on
DTDs/XSDs and it shows how some specificities of the XML data model can be exploited in the design of a matcher.
Second, it describes how some recent results in the area of XML Matchers can be used in innovative and emerging
applications in the broad area of Data Management. In particular, it focuses on the task of {\em clustering schemas},
i.e., on automatically grouping heterogeneous DTDs/XSDs. Schema clustering is of the utmost relevance at the Web scale
because the number, size and complexity of available data sources are typically huge and, therefore, it is impractical
to manually (or semi-automatically) classify schemas into pre-defined domains. Schema clustering techniques allow data
sources on the Web to be organized into homogeneous groups; after this task, it is possible to adopt existing
approaches to integrate the schemas belonging to same group. This yields practical advantages: for instance, once a
user submits a query $q$, it is possible to rank available domains on the basis of their relevance to $q$. In this way,
it is possible to achieve a twofold benefit: the answers a use get back are more precise (because only those data
sources which are likely to generate sound and correct results are selected), and the time required for processing
queries is reduced (because irrelevant data sources are not contacted).

In the context of Semantic Web, Ehrig \cite{Ehrig06} focused on the problem of Ontology Alignment/Ontology Matching
(which strongly resembles the Schema Matching problem) and depicted the process of aligning ontologies as a six-step
process.

Ontology Matching (as well as its relationship with Schema Matching) has been also reviewed in \cite{ShEu05} and in the
subsequent book \cite{EuSh07}. A very recent update on the state-of-the-art in Ontology Matching can be found in
\cite{ShEu13}. Ontology matching differs from the matching of E/R and relational schemas. Indeed, first ontologies
provide a high flexibility level because they offer a large number of primitives (e.g., cardinality constraints,
disjoint classes, hierarchical organization of concepts, and so on) not available in E/R and relational schemas.
Secondly, an E/R (relational) schema is usually designed for modeling a specific piece of reality and, in general, it
is hard to re-use an existing E/R (resp., relational) schema in other contexts; by contrast, ontologies are by
definition reusable and sharable. Finally, the design of an E/R (resp., relational) schema is generally delegated to
human experts who generally share the same vocabulary. Instead, the design of an ontology is becoming more and more a
decentralized effort in which multiple independent contributors are in charge of designing and populating the ontology
itself. From this discussion it emerges that the task of matching DTDs/XSDs lies between the matching of E/R (resp.,
relational) schemas and that of ontologies: as in the ontology matching, the hierarchical structure of DTDs/XSDs can
convey semantics (see Section \ref{sec:matching-xml}) and can be advantageously used in matching discovery. A big
difference between the matching of DTDs/XSDs and that of ontologies is that the latters can be defined as a set of
logical axioms useful to define the semantics of data; this information is not available in the matching of DTDs/XSDs.

A novel, definitely relevant, trend in Schema Matching research regards the problem of managing uncertainty in Schema
Matching process \cite{Gal11}. An excellent review of uncertainty in Schema Matching is proposed by Gal \cite{Gal11}.
In this book, the author presents a framework to classify the various aspects of uncertainty. The book provides also
several alternative representations of Schema Matching uncertainty and discusses in depth some strategies that have
been recently proposed to deal with this issue.

Our work focuses on a narrow sub-area of Schema Matching, namely Schema Matching of XML sources. As far as this
sub-area is concerned, it significantly extends published surveys. In particular, the main contributions of our survey
can be summarized as follows:

\begin{enumerate}

\item Most of the existing work is agnostic of the data model. By contrast, our survey focuses on XML-specific
    Schema Matching techniques. We do not consider well-known techniques that have been developed for some data
    models and have been subsequently reused on XML, but we discuss in detail the specificities of DTD/XSD and
     how they influence the Schema Matching problem (see Section \ref{sec:matching-xml}).

\item We discuss in detail potential application fields benefiting from XML-specific Schema Matching. Due to the
    widespread adoption of XML in the business domain, we pay a special attention to commercial applications.

\item We discuss the problem of uncertainty management in the context of XML-specific Schema Matching. In fact, we
    recognize the enormous impact that uncertainty can have in a wide range of real-world applications based on
    Schema Matching. While significant research efforts have been done in the field of uncertainty management,
    there are very few approaches dealing with uncertainty in XML-specific Schema Matching.

\end{enumerate}

\section{Schema Matching algorithms}
\label{sec:schema-matching-review}

\subsection{Basic aspects of Schema Matching algorithms}
\label{sub:basic}

{\em Schema Matching} aims at finding relationships between elements of two schemas \cite{RaBe01}. In the Semantic Web
literature, the Schema Matching problem is also known as {\em Ontology Alignment} or {\em Ontology Matching}. We refer
the reader to \cite{ShEu05} for a deep discussion about the main differences between the Schema Matching and the
Ontology Alignment problems.

A schema is a structure encoded in a formal language that describes a piece of reality; examples of schemas are
relational schemas, E/R diagrams, Ontologies, DTD, XSD, and so on. The schema describing a data source is often known
as the {\em intensional component} of that data source, whereas the set of instances associated with it is called {\em
extensional component}. A schema generally consists of a set of entities (called {\em schema elements} or {\em schema
components}), representing real-world objects, a set of relationships, specifying connections among schema elements,
and a set of constraints. In the following we will denote as $\mathcal{E}(S)$ the set of schema elements associated
with a schema $S$. For instance, in case of a relational database, $\mathcal{E}(S)$ coincides with a set of tables,
whereas, in case of XSD, $\mathcal{E}(S)$ is the set of elements, attributes and complex types in $S$.

The relationships derived by the Schema Matching process are often known as {\em semantic matchings} or, equivalently,
as {\em mappings} \cite{ShEu05} or {\em interschema properties} \cite{BaLeNa86,Ursino-TKDE1}. An algorithm designed to
find semantic matchings is often called, in short, {\em Matcher} \cite{RaBe01}.

A more formal definition of semantic matching is given in \cite{ShEu05}. We report it below:

\begin{definition}({\em Semantic Matching}).
\label{def:matching} {\em Given two Schemas $S_1$ and $S_2$, a semantic matching is a tuple $\langle id, u, v,
\overline{c}, \overline{r}\rangle$ where: {\em (i)} $id$ is the identifier of the semantic matching; {\em (ii)} $u$ and
$v$ are two schema elements belonging to $S_1$ and $S_2$, respectively; {\em (iii)} $\overline{c}$ is a confidence
measure (generally ranging in the [0,1] real interval) stating the strength of the relationship between $u$ and $v$ and
{\em (iv)} $\overline{r}$ is a relation holding between $u$ and $v$. }$\punto$
\end{definition}

The most common specifications of $\overline{r}$ are {\em synonymy}, {\em homonymy}, {\em hyponymy} and {\em
hyperonymy}. In detail, given two elements $u \in S_1$ and $v \in S_2$ we say that:

\begin{itemize}

	\item $u$ and $v$ are {\em synonyms} if they have the same meaning, even if they could have different names;

	\item $u$ and $v$ are {\em homonyms} if they have different meanings, even if they have the same name;

	\item $u$ is a \emph{hyponym} of $v$ (which, in its turn, is a \emph{hyperonym} of $u$) if $u$ has a more
specific meaning than $v$ (e.g., $u$  may be the element ``PhD student'' whereas $v$ may be the element
``student'').

\end{itemize}

As pointed out in Section \ref{sec:related-work}, one of the first attempts to classify Schema Matching approaches was
proposed in \cite{RaBe01}. Here, we summarize some of the classification criteria introduced therein.

\subsubsection{Schema-level Matchers vs. instance-level Matchers}
\label{subsub:schemainstance}

{\em Schema-level Matchers} assume that each available data source is provided with a schema representing it
\cite{ShEu05,RaBe01}. The schema of a data source represents a rich body of information, useful to carry out matching
activities. For instance, a schema allows for the extraction of the {\em name} of its elements, of their {\em data
types} and of some of their {\em constraints} (e.g., cardinality constraints).

\begin{figure}
\centering
\includegraphics[width=11cm]{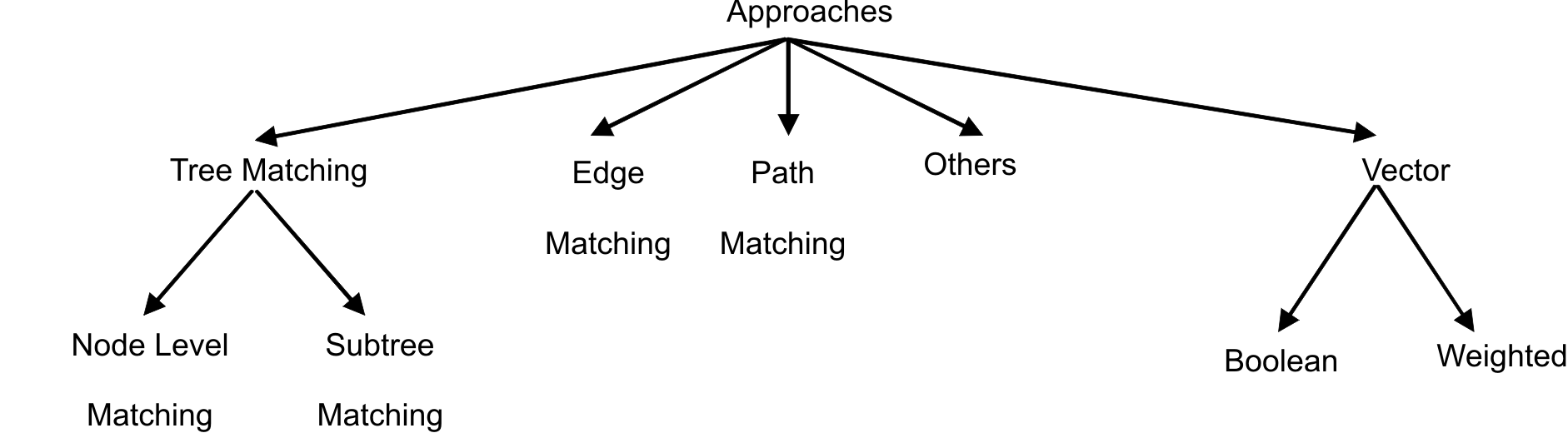}
\caption{A graphical classification of approaches to computing similarities between XML documents}
\label{fig:XMLDocumentMatching}
\end{figure}

{\em Instance-level Matchers} analyze the extensional component of a data source to infer semantic matchings. They are
generally {\em very accurate} because they look at the {\em actual content} of the involved sources. However, they are
computationally expensive since the amount of data they need to process can be large. Strictly related to
instance-level matchers, matching-related approaches for XML documents have been largely investigated in the
literature. Indeed, this problem has attracted the interest of many researchers mainly in the areas of Database Systems
and Information Retrieval
\cite{Dalamagas*06,CoAbMa02,NiJa02,Costa*04,AnMaTs08,Buttler04,RaMoSu06,Flesca*05,ThScWe03,YaChCh05}. A graphical
classification of these approaches is reported in Figure \ref{fig:XMLDocumentMatching}. In the following, we overview
the key ideas and concepts underlying each of these main approach categories. {\em Tree matching approaches}
\cite{Dalamagas*06,CoAbMa02,NiJa02} model XML documents as labeled trees. Often, they use dynamic programming
techniques to find the distance of two trees. In particular, the main idea of these approaches is that the distance of
two trees coincides with the minimum number of operations (called edit operations) capable of transforming the former
tree into the latter one. {\em Edge matching approaches} \cite{Costa*04,AnMaTs08} require to transform XML documents
into directed graphs and to compute the similarity of two documents based on their common edges. {\em Path matching
approaches} \cite{Buttler04,RaMoSu06} model an XML document as a set of paths, each starting from the root of the
document and ending in a leaf node; therefore, the similarity between two XML documents can be computed based on the
sets of paths associated with them. {\em Vector based approaches} \cite{ThScWe03,YaChCh05} suggest to map XML documents
onto vectors of an abstract n-dimensional feature space and to compute the matching between two documents as a suitable
distance between the corresponding vectors.

In real scenarios, instance-level Matchers are often used, along with schema-level ones, to filter out {\em false
matchings} and to learn similarities among schema elements on the basis of the similarity degrees of the corresponding
instances. Matchers adopting this strategy are known as {\em  hybrid Matchers} \cite{RaBe01,Kementsietsidis09}. The
overall result is an increase of the accuracy of discovered matchings. However, as in the case of pure instance-level
Matchers, the amount of data to process is usually very large, and, therefore, the computational cost can be
significantly high. Hybrid Matchers often rely on {\em machine learning algorithms}, like {\em Self-Organizing Maps}
\cite{Kohonen88}, or on a combination of classifiers \cite{DoDoHa01,Doan*02}.

\subsubsection{Simple Matchers vs. complex Matchers}
\label{subsub:simple}

Matchers can be also classified on the basis of the cardinalities of the matchings they are able to find \cite{RaBe01}.
To better illustrate this concept, let us consider two schemas $S_1$ and $S_2$. The following kinds of Matchers can be
defined:

\begin{itemize}

	\item {\em Simple Matchers or 1:1 Matchers}. They aim at finding pairs of elements $\langle u,v \rangle$ such
that $u \in S_1$, $v \in S_2$ and a semantic matching exists between them. Most of these approaches derive
synonymies, but some of them consider also hyponymies/hyperonimies \cite{Ursino-Informatica1,Gal*05}.

    \item  {\em Complex Matchers or m:n Matchers}. They find pairs of the form $\langle G_1, G_2 \rangle$, being
        $G_1$ and $G_2$ two groups of elements extracted from $S_1$ and $S_2$, respectively. For instance, $G_1$
        could be an element ``address'', whereas $G_2$ could be a pair of elements $\langle$``street'',
        ``zip''$\rangle$. Such a matching indicates that the group of elements ``street'' and ``zip'' of $S_2$ is
        semantically similar to the element ``address'' of $S_1$. Complex Matchers are also known as {\em
        sub-schema similarities} \cite{Ursino-COOPIS2004}.

\end{itemize}

Most of Schema Matching algorithms return only {\em 1:1} matchings \cite{DoDoHa01,DoRa02,GiYaSh07}. However, there are
few examples of systems capable of handling complex matchings, such as iMAP \cite{Dhamankar*04}, DCM \cite{HeCh03},
INDIGO \cite{IdVa07} and XIKE \cite{Ursino-COOPIS2004}. In iMAP the computation of complex matchings is seen as a {\em
search problem}, and a set of searchers is used to explore the space of all the possible groups of schema elements
candidate to form a complex matching. In order to reduce the computational complexity, the search space is pruned by
taking domain knowledge into account. DCM hypothesizes the presence, for each application domain, of a {\em hidden
schema} which acts as a unified generative model describing how schemas are generated from a finite vocabulary. INDIGO
uses a variety of techniques, like linguistic Matchers and the {\tt WHIRL} algorithm \cite{CoHi98}, a $k$-nearest
neighbor classifier developed in the context of text classification. XIKE will be extensively discussed in the next
sections because it has been explicitly designed to deal with XSD.

Matchers often use algorithms from Natural Language Processing to carry out a pre-processing step
\cite{GiYaSh07,MaBeRa01}. For instance, some popular pre-processing activities are {\em tokenization} (i.e., the names
of the entities are parsed, and symbols like punctuation, blank characters or digits are detected), {\em elimination}
(i.e., some tokens, like prepositions or articles, are filtered out), and {\em stemming} (i.e., derived or inflected
words are reduced to their stem, or root).

Most of the existing Matchers rely on the use of {\em external knowledge bases}. The simplest examples of these
knowledge bases are dictionaries or thesauri. A popular thesaurus is {\em WordNet}, a lexical database for English
language developed at the University of Princeton \cite{Miller95}. WordNet groups English words into sets of synonyms
(called {\em synsets}) and, then, provides a plenty of facilities for Schema Matching purposes. WordNet has been used
in the {\em S-Match} system \cite{GiYaSh07,ShGiYa10} as a {\em background knowledge source} to return semantic
relationships between element names. Other approaches suggest to use Wikipedia as an alternative or in combination with
WordNet. Knowledge bases like Wikipedia attracted the interest of researchers investigating the management of data
sources from an instance-level perspective \cite{Tagarelli13}. For instance, \cite{PoSt07} uses Wikipedia for computing
semantic relatedness and, on some datasets, the results achieved by means of Wikipedia are more precise than those
obtained by means of WordNet; other approaches \cite{BuPa06} show the effectiveness of Wikipedia in word sense
disambiguation. Wikipedia has been coupled with WordNet in the context of YAGO (Yet Another Great Ontology) ontology
\cite{SuKaWe07}. Yago contains more than 1 millions entities and 5 millions facts: entities represent, for instance,
people, organizations, products, and so on, whereas facts link entities (e.g., a fact could be {\em ``Albert Einstein
won the Nobel Prize''}). Facts are automatically extracted from Wikipedia which offers quite an extensive coverage of
human knowledge. Wikipedia supplies also some {\em category pages} to organize facts and entities into a hierarchy,
but, as observed in \cite{SuKaWe07}, such a hierarchy is rather imprecise and, therefore, it is barely useful for
ontological purposes. WordNet provides a rich and clean hierarchy consisting of thousands of concepts. Yago solves the
non-trivial task of creating a link between WordNet concepts and entities/facts in Wikipedia. The result is a
knowledge-base which significantly extends WordNet in quantity (because it increases the number of facts managed by
WordNet by more than one magnitude order) and in quality (because it adds knowledge about individuals/organizations by
specifying their semantic relationships).

A new research direction in Schema Matching field focuses on the design of Matchers capable of handling large schemas
in an efficient fashion without sacrificing the accuracy of the matching process. A further, promising research avenue
aims at exploring how user feedbacks can be exploited to improve the accuracy of the matching process. In the next
subsections we discuss the corresponding techniques.

\subsection{Efficient computation of semantic matching}
\label{sub:efficient}

Large size schemas are becoming more and more frequent especially in the business domain \cite{DoRa07,RaDoMa04,Rahm11}.
This implies that the size  of the space of potential matchings to explore significantly grows. This poses relevant
computational challenges and may negatively impact on the accuracy of a Matcher \cite{DoRa07}.

Therefore, it is not surprising that several authors were involved in designing efficient Matchers capable of handling
large schemas. Some of these approaches rely on the idea of filtering out all the pairs of elements which are not
likely to form a matching \cite{EhSt04}. Other ones (called {\em partition-based}) use a {\em divide et impera}
strategy to perform matching. In these approaches, a schema or an ontology is fragmented into smaller portions and a
Matcher is recursively applied on each of these partitions. Partitioning provides the chance of parallelizing Schema
Matching tasks. A relevant example of partition-based approaches is the COMA++ system \cite{DoRa07}.

More recently, some authors \cite{Rahm11} borrowed ideas from the {\em entity resolution} research field to reduce the
size of the search space of potential matchings. For this purpose, they suggest to cluster some selected attributes;
such a procedure is called {\em blocking} \cite{ElIpVe07}. Some authors proposed to use the MapReduce framework to
perform blocking \cite{KoThRa11}; for instance, a prototype, called {\em Dedoop} (Deduplication with Hadoop), was
implemented to perform entity resolution in large datasets \cite{KoThRa12}.

\subsection{Incorporating user behavior analysis in Schema Matching}
\label{sub:Incorporating-User-Behaviors}

Some recent approaches propose to incorporate information describing user behaviors in the computation of semantic
matchings. These approaches assume that users interact with a group of information systems by submitting queries over
time to search for information of their interest. User queries can be analyzed to find groups of frequent attributes.
The main assumption behind these approaches is that if two or more attributes frequently co-occur in user queries then
they are likely to form a semantic matching.

The first approach belonging to this category was proposed in \cite{ElOuEl08}. It aims at finding pairs/groups of
attributes which frequently co-occur in user queries. Further features are also considered to establish the strength of
the association between two or more attributes (e.g., how frequently they co-occur in joins or as arguments of
aggregate functions). Several scoring functions are implemented to assess the similarity score of a pair (resp., group)
of attributes. Finally, a genetic algorithm is exploited to find the groups of attributes having the highest similarity
scores.

The approach of \cite{ElOuEl08} works on relational databases. An extension of it to the Web scenario is proposed in
the HAMSTER system \cite{NaBe09}, which analyzes the query log of a search engine to discover mappings between the
concepts of two schemas.

The approaches of this category are quite interesting because the analysis of user behaviors can highlight forms of
correlations between elements of different schemas which cannot be revealed by traditional Schema Matching techniques.

A further advantage is that, in some cases, no knowledge of the intensional component of a data source is available
and, at the same time, the knowledge of the extensional component could be partial or of bad quality. These facts could
make the final quality of the matching process poor. In this case the analysis of user behaviors in accessing these
data sources could be precious to better understand the corresponding structure and, ultimately, to derive semantic
matchings.

A main problem of these approaches is that they require data about user past behaviors. In many cases these data are
not available since they are handled by search providers and, due to privacy and commercial limitations, they cannot be
disclosed. A further problem is that the approaches belonging to this category are biased by search terms frequently
exploited by users. In fact, if a term is adopted in many user queries, then a large body of information on it is
available, and this helps to find semantic matching. By contrast, if a term is rarely used, the corresponding available
information is poor, and this could have a negative impact on the results of the matching activity.

\section{The role of DTD and XSD in XML Matchers}
\label{sec:matching-xml}

In this section we describe to what extent the specificities of DTD/XSD impact on the Schema Matching process.
Throughout the paper, for sake of simplicity, when we say ``a DTD'' (resp., ``an XSD'') we mean the intensional
component of an XML data source encoded by means of the DTD (resp., XSD) language. Similarly, when we say ``matching of
two DTDs (resp., XSDs)'' we mean the matching of the intensional components of two XML data sources encoded by means of
the DTD (resp., XSD) language.

In the following subsections we discuss how the features of DTDs/XSDs may impact on the semantics of schemas encoded by
means of these languages (Section \ref{sub:xml-data-model-matching}). After that, we show that XSD offers some advanced
features not present in DTD. These features can provide further insights to find semantic matchings, but their usage is
far from trivial. In detail, we first discuss how to use constraints on data types (Section
\ref{sub:constraint-data-type}), and, then, constraints on cardinalities (Section \ref{sub:cardinalities}). Finally, we
focus on the practice of reusing XSDs, or part of them (Section \ref{sub:reuse}).

\subsection{Specificities of DTDs/XSDs in XML Matching}
\label{sub:xml-data-model-matching}

Almost all the approaches described and classified in Section \ref{sec:schema-matching-review} do not prescribe a
specific data model. In most cases, these approaches handle database schemas encoded in a generic data model (e.g., a
pair of E/R diagrams \cite{Ursino-TKDE1,MeGaRa02}, an XSD and a relational schema \cite{MeGaRa02}, and so on) and
convert them into an internal model (e.g., a graph or an object-oriented schema) which is, then, used to derive
semantic matchings.

However, the features of the data model adopted to design a database may have a great impact on the semantics embedded
in the corresponding schema and, consequently, on the derivation of semantic matchings.

This means that we could use, for instance, E/R diagrams and DTDs/XSDs to represent the same piece of reality, but we
may expect that the semantics captured by E/R diagrams may differ from that encoded in DTDs/XSDs, and vice versa.

To better point out this fact, we consider a running example that we will use throughout this section. It deals with a
{\em University Library}.

\begin{example}
\label{ex:intro}
{\em Let us consider a simple University Library. Users are University Students, Professors, University Staff and Foreign
Visitors. The Library offers several types of publications, like Books, Articles (i.e., publications on scientific
journals), Papers (i.e., publications on conference proceedings), Abstracts, Government Publication, and so on. In this
library, there are different types of employees, with different kinds of qualification (secretary, stock manager,
shelver, and so on).
Assume we want to model only {\em book authorship} relationship. We use an E/R diagram and, then, an
XSD\footnote{Analogous considerations would hold if the selected data model was DTD, instead of XSD.}.
The E/R diagram describing the authorship relationship is reported in Figure \ref{fig:ERdiagram}: It shows two
entities, namely {\em Book} and {\em Author}, joined by the relationship {\em Writes}.

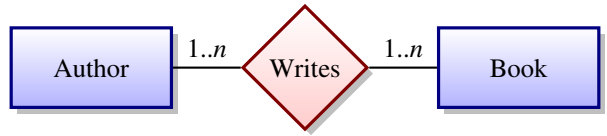
\begin{figure}[h]
\centering
\begin{tikzpicture}[node distance=8em, every edge/.style={link}]
\node[entity] (author) {Author};
\node[relationship] (writes) [right of=author] {Writes}
edge node[auto,swap] {1..$n$} (author);
\node[entity] (Book) [right of=writes] {Book}
edge node[auto,swap] {1..$n$} (writes);
\end{tikzpicture}
\caption{The E/R diagram describing {\em Book}(s) and {\em Author}(s).}
\label{fig:ERdiagram}
\end{figure}

There are different options to translate this diagram into an XSD \cite{PiQu05}. A first one requires the mapping of
the entities {\em Book} and {\em Author} onto two elements of an XSD and the introduction of a father/child
relationship between them. In particular, {\em Author} (resp., {\em Book}) can be set as the father element and {\em
Book} (resp., {\em Author}) as the child one. To graphically represent the obtained XSD, we will use a tree-based
diagram in which each schema element corresponds to a node, whereas an edge links an element with one of its
sub-elements. The tree diagram representing the first (resp., second) translation option is reported in Figure
\ref{fig:xmlschemabookstore}(A) (resp., Figure \ref{fig:xmlschemabookstore}(B)). Another option, which is graphically
reported in Figure \ref{fig:xmlschemabookstore}(C), prescribes the introduction of a third element {\em Writes}. In
this case, we need to use the {\tt KEYREF} construct between the attributes of {\em Writes} and the key attributes of
{\em Book} and {\em Author}. Each of these alternatives reflects the personal standpoint of a human designer. In
practical cases, the best option among the available alternatives is usually the one avoiding redundancies
\cite{PiQu05}.

\begin{figure}[h]
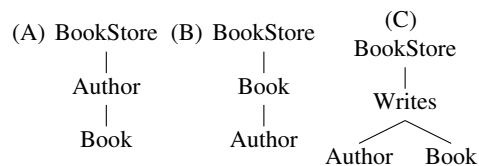
%
	\centering
	\small
	
	\begin{minipage}{2cm}
		\begin{center}
        \small (A)
        \smallskip
		\Tree [.{BookStore} [.{Author} [.{Book} ]]]		
		\end{center}		
	\end{minipage}
	\begin{minipage}{2cm}
		\begin{center}
		\small (B)
        \smallskip
        \Tree [.{BookStore} [.{Book} [.{Author} ]]]
	    \end{center}
	\end{minipage}
	\begin{minipage}{2cm}
		\begin{center}
		\small (C)
        \Tree [.{BookStore} [.{Writes} [.{Author} ] [.{Book} ]]]		
		\end{center}		
	\end{minipage}
	
	\caption{Three XSDs describing {\em Books} and {\em Authors}.}%
	\label{fig:xmlschemabookstore}%
\end{figure}
}
$\punto$
\end{example}

On the other hand, it is straightforward to observe that also the translation from XSDs to other data models, like E/R
diagrams, relational schemas or object-oriented data models, incurs in some difficulties. As previously pointed out,
this depends on the fact that XSD (as well as DTD) provides some features not present in E/R diagrams and in
object-oriented data models. DTD and XSD, in fact, allow for a hierarchical organization of information, and this
feature is not present in E/R diagrams or relational schemas. In addition, the sibling order is a main XML
specification; this implies that the order at which a child of an element occurs is relevant. This characteristic is
missing in E/R, relational and object oriented data models.

These features may have a huge impact on the performance of a Matcher: For instance, the role of the hierarchical
structure of a DTD in conveying semantics was investigated in the {\sf LSD} system \cite{DoDoHa01}. Here, the DTDs to
match are considered in conjunction with the associated documents. In particular, an XML document is {\em serialized},
i.e. it is mapped onto a list of pairs of the form $\langle \mathtt{tag\_name}, \mathtt{content} \rangle$, where
$\mathtt{tag\_name}$ specifies the name of an element and $\mathtt{content}$ specifies its value. To better clarify
this concept, let us consider, again, Example \ref{ex:intro} and, in particular, an element {\tt employee} having {\tt
name}, {\tt surname} and {\tt phone} as attributes.

Let us consider an instance of this element:

{\scriptsize
\begin{lstlisting}
         <employee>
                 <name> John </name>
                 <surname> Smith </surname>
                 <phone> (001) 514 201 </phone>
         </employee>
\end{lstlisting}
}

This fragment is converted in the list:

\begin{center}
{\small

$\langle \mathtt{Name}, John \rangle, \langle \mathtt{Surname}, Smith \rangle, \langle \mathtt{Phone},$ {\em (001) 514
201} $\rangle$

}
\end{center}

Serialization leads to make the hierarchical structure of an XML document not explicit.

Experiments show that the serialization of an XML document may increase the number of false negatives, and this
negatively impacts on the accuracy of the corresponding Matchers. In order to overcome this drawback, {\sf LSD}
explores also a {\em more refined encoding} of an XML document which considers the {\em nesting level} of each data
instance and yields a meaningful reduction of the number of false negatives (and, then, an increase of the Matcher's
accuracy). The case examined above, in the whole, highlights that the hierarchical structure of a DTD is somewhat
related to its semantics, and it has to be properly considered in the design of a Matcher.

In line with the findings of LSD, Gal {\em et al.} \cite{Gal*05,MoGaJa01} studied how to use the structural information
provided by a specific data model (like HTML or XML) to improve the accuracy of Schema Matching process. In particular,
they propose to exploit derived matchings to help users in seeking information of their interest. This is a nice
application of Schema Matching in the context of Information Retrieval \cite{MoGaJa01}. In detail, in \cite{MoGaJa01},
the hierarchical structure of a group of HTML pages is analyzed to extract an ontology which is, subsequently,
exploited to refine user queries and, then, to identify pages of interest for the users.

A further, relevant contribution is discussed in \cite{Gal*05}. In this paper, the authors show that structural
information is useful to identify links among concepts present, for instance, in an XML/HTML document, and this
information could be considered in the Schema Matching process. The approach of \cite{Gal*05} was applied to find
matchings between pairs of Web sources which were then exploited to help users in the search of information on the Web.
Experiments carried out in real-life scenarios (like car rental or airline Web sites) showed that semantic matchings
were effective in enhancing the precision of Web search. This is a further confirmation that structural information
provides a valuable contribution to the Schema Matching process.

\subsection{Constraints on data types}
\label{sub:constraint-data-type}

Both DTD and XSD provide a rich set of primitive and built-in data types. For instance, XSD offers 43 built-in simple
types, like {\tt string}, {\tt integer}, {\tt float}, {\tt boolean}, {\tt time} and {\tt date}, which can be used for
declaring elements and attributes \cite{XSD12}.

The compatibility of the data types associated with two elements is a useful indicator to assess whether the last ones
actually form a semantic matching or not. Constraints on data types can be {\em hard} (if they cannot be violated in
any case) or {\em soft} (if they can be relaxed). For instance, a hard constraint could specify that a {\tt date}
cannot be converted into a {\tt real}, whereas a soft constraint could specify that an {\tt integer} can be converted
into a {\tt longinteger} in some cases. Constraints on data types are relevant to detect false positives: Given a pair
of schema elements $u$ and $v$, we can conclude that they do not form a semantic matching if there is a hard constraint
stating that the data type of $u$ cannot match with the one of $v$. The reasoning above shows that constraints on data
types are straightforward to manage if the involved types are primitive or, more in general, if they can be solved by
means of compatibility tables \cite{AlNaSa10,Ursino-IS4,MaBeRa01}. An example of compatibility table is reported in
Table \ref{tab:compatibility-table}.

\begin{table}
\centering

{\scriptsize
\begin{tabular}{||c|c|c||}
\hline \hline
{\em Type 1} & {\em Type 2} & {\em Compatibility Degree} \\
\hline \hline
{\tt string} & {\tt string} & 1.0 \\
\hline
{\tt string} & {\tt decimal} & 0.2 \\
\hline
{\tt decimal} & {\tt float} &0.8 \\
\hline
{\tt float} & {\tt float}  &1.0 \\
\hline
{\tt float} & {\tt integer} & 0.8 \\
\hline
{\tt integer} & {\tt short} & 0.8 \\
\hline \hline
\end{tabular}
}

\caption{A table to check data type compatibility.} \label{tab:compatibility-table}
\end{table}

The first and second column of Table \ref{tab:compatibility-table} reports two data type (like {\tt float} or {\tt
string}), whereas the third column reports the corresponding compatibility degree. This last coefficient ranges from 0
to 1, and the higher its value the higher the level of compatibility of the corresponding data types.

Even simple types defined by users can be easily managed, because they can be usually associated with a built-in simple
type.

The management of data types becomes much harder in case of XSD. In fact, this data model allows human designers to
define their own data types (called {\em complex types}) by assembling simple elements and/or attributes.

Furthermore, designers are allowed to create new data types starting from an existing one (called {\em base type}).
Such a process implicitly induces a relationship between a base type and the new types derived from it. For instance, a
designer may declare a complex type by extending an already existing one by means of the $\mathtt{<xs:extension>}$
construct. In this way, a derived complex type contains all the elements of the base type plus additional ones,
specific of the new type. Such a construct can be, therefore, intended as a {\em generalization}. In an analogous
fashion, a designer may set some restrictions on an base type, and this means that the values associated with a derived
type are a subset of the ones that can be assumed by the base type. In this way, the $\mathtt{<xs:restriction>}$
construct can be used to implement a {\em specialization} relationship. Finally, the two constructors
$\mathtt{<xs:sequence}>$ and $\mathtt{<xs:all>}$ implement the concept of {\em aggregation}, i.e. they specify that a
complex type consists of many sub-elements or, equivalently, that each sub-element is {\em part of} the complex type.
If the $\mathtt{<xs:sequence>}$ construct is adopted, sub-elements must appear in the same order as they have been
declared. Vice versa, if $\mathtt{<xs:all>}$ is used, sub-elements may appear in any order.

The constructs outlined above contribute to specifying the semantics of an element of an XSD in relation to the one of
some already defined elements and, therefore, can be exploited by a Matcher.

By contrast, the complexity underlying the management of compatibility between complex types is, in principle,
unlimited. In fact, a designer can decide to repeatedly assembling simple elements, attributes and already defined
complex elements, to construct complicated complex elements. In such a case, the matching of two complex elements can
become as much hard as the matching of two XSDs \cite{RaDoMa04}.

\subsection{Constraints on cardinalities}
\label{sub:cardinalities}

A further constraint regards the cardinality associated with the instance of an element in a document.

In case of DTD, a quantifier is used to specify the number of occurrences of an element. Allowed quantifiers are: {\tt
+} (indicating that there must be one or more occurrences of the element), {\tt *} (denoting that zero or more
occurrences are allowed), {\tt ?} (indicating that no more than one occurrence is allowed), and {\tt None}.

Analogously to data type compatibility, it is possible to check whether the cardinalities of two schema elements differ or not and, if the differences are irreconcilable, the corresponding candidate matchings can be discarded. Cardinality constraints can be managed by means of suitable tables \cite{Lee*02,WoMlDo10}. An example of this table (originally proposed in \cite{Lee*02}) is that reported in Table \ref{tab:cardinality}.

\begin{table}
\centering

{\scriptsize
\begin{tabular}{||c|c|c|c|c||}
\hline \hline
     & {\tt *} & {\tt +} & {\tt ?} & {\tt None} \\
\hline \hline
{\tt *}    & 1 & 0.9 & 0.7 & 0.7\\
\hline
{\tt +}    & 0.9 & 1 &0.7  & 0.7\\
\hline
{\tt ?}    & 0.7 & 0.7 & 1 & 0.8\\
\hline
{\tt None} & 0.7 & 0.7 &0.8 & 1\\
\hline \hline
\end{tabular}
}

\caption{A table to check data cardinality constraints.}
\label{tab:cardinality}
\end{table}

In XSD, the cardinalities of elements are declared by means of the attributes {\tt minOccurs} and {\tt maxOccurs}, and
this information can be used in the semantic matching discovery. Nayak and Tran \cite{NaTr07} adapted the different
methods for handling cardinalities in DTD/XSD; the conversion schemas proposed by them is reported in Table
\ref{tab:equivalence}.

\begin{table}[t]
\centering

{\scriptsize
\begin{tabular}{||c|c|c||}
\hline \hline
{\em DTD Cardinality Symbol} & {\em Rule} & {\em XDS occurrences}\\
\hline  \hline
{\tt +} & is equivalent to &  ({\tt minOccurs = 1}; {\tt maxOccurs = unbounded}) \\
\hline
{\tt *} & is equivalent to &  ({\tt minOccurs = 0}; {\tt maxOccurs = unbounded}) \\
\hline
{\tt ?} & is equivalent to &  ({\tt minOccurs = 0}; {\tt maxOccurs = 1}) \\
\hline
{\tt None} & is equivalent to &  ({\tt minOccurs = 1}; {\tt maxOccurs = 1}) \\
\hline \hline
\end{tabular}
}

\caption{Equivalence table between cardinality constraints in DTD/XSD.} \label{tab:equivalence}

\end{table}

\subsection{Reuse of Schema Elements}
\label{sub:reuse}

XSD offers several methods for reusing already defined schema elements. In the following we focus on {\em shared
components} and {\em distributed name spaces}, and we illustrate their impact on Schema Matching.

As for shared components, we observe that in XSD an element is called {\em global} if it is defined as child of the
$\mathtt{schema}$ root element. Global elements are the only ones that can be referenced, and, therefore, they can be
reused in the definition of new data types. To clarify this concept, let us consider, again, the University Library
scenario described in Example \ref{ex:intro}.

\begin{example}
\label{ex:shared}
{\em Suppose that in the University Library there are (at least) three kinds of item, namely {\tt Book}(s), {\tt
Art\-icle}(s) and {\tt Paper}(s). We could define three different types for representing them; alternatively, we may
define a global element {\tt Publication}, reporting information common to books, papers and articles:

{\scriptsize
\begin{lstlisting}
      <xsd:complexType name="Publication">
              <xsd:sequence>
                  <xsd:element name="Title" type="Title"/>
                  <xsd:element name="Author" type="Name"/>
                  <xsd:element name="Editor" type="Name"/>
              </xsd:sequence>
      </xsd:complexType>
\end{lstlisting}
}

\noindent and three elements {\tt Book}, {\tt Article} and {\tt Paper}, defined as follows:

{\scriptsize
\begin{lstlisting}
      <xsd:element name="Book" type="Publication"/>
      <xsd:element name="Article" type="Publication"/>
      <xsd:element name="Paper" type="Publication"/>
\end{lstlisting}}
}$\punto$
\end{example}

Global elements are often called {\em shared components} \cite{RaDoMa04}.

Thanks to global elements, a designer can avoid to unnecessarily define the same object multiple times; indeed, she can
define it once and reuse it as many times she wants. Of course, some designers may opt for a heavy reuse of global
elements, whereas others (especially in case of small schemas) may prefer to not use them. If a designer does not use
shared components, the resulting XSD can be graphically described as a tree in which nodes represent schema elements
and edges model relationships between elements and sub-elements. By contrast, if she adopts shared components, it is
easy to argue that the XSD may appear as a graph with loops.

To better clarify this concept, we will provide a simple example.

\begin{example}
\label{ex:cycles-in-XSD}
{\em Consider once again the University Library introduced in Example \ref{ex:intro}. A scientific publication could be
accompanied by additional material, like a software prototype and the datasets used in the experimental evaluation of
the approach described in it. In the XSD describing the University Library, we can model this scenario by adding an
element {\tt Additional\_Material} as a {\em global element}. Suppose that this element specifies a {\tt URL} (pointing
to a Web site containing that material) as well as a description of the corresponding material. This description could
specify, for instance, that the additional material is a collection of software programs (used to run experiments
presented in the paper), one or more datasets (used to test the proposed approach) and technical documentations
(specifying, for instance, the policy adopted for protecting personal data). Assume, now, that the element {\tt
Publication} has been defined as a global element too.

Let us consider, now, a new element {\tt Demo\_Paper}, describing a paper presented in the experimental track of a
conference or a workshop. We could define this element from scratch, by specifying its title, its number of pages, a
Web link to the software prototype described in the paper, a link to the dataset used in the corresponding experiments,
and so on. Alternatively, we can define it by reusing the global elements {\tt Paper} and {\tt Additional\_Material} we
have at our disposal. This leads to an XSD fragment which can be graphically represented as in Figure
\ref{fig:xsd-shared-components}. Here, we reported the name of schema elements and we drew an arrow from a global
element $a$ to an element $b$ to specify that $b$ references $a$ (so, for instance, {\tt Demo\_Paper} references {\tt
Additional\_Material}).

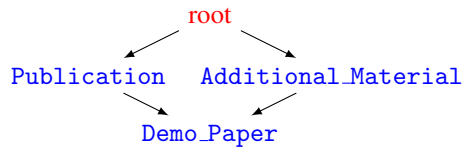
\begin{figure}[th]%
	\centering 	
	\tikzstyle{node}=[circle,fill=black!25,minimum size=20pt,inner sep=0pt]
	\tikzstyle{edge} = [draw,thick,-]
   \tikzstyle{weight} = [font=\footnotesize] 	
	\begin{tikzpicture}
  [scale=.8,auto=left]
  \node (a) at (0,0) {{\color{red}root}};
  \node (b) at (2,-1) {{\color{blue}{\tt Additional\_Material}}};
  \node (c) at (-2,-1) {{\color{blue}{\tt Publication}}};
  \node (d) at (0,-2) {{\color{blue}{\tt Demo\_Paper}}};

  \foreach \from / \to / \weight in {a/b/{\color{red}},a/c/{\color{red}},b/d/{\color{red}}, c/d/{\color{red}}}
     \draw[->, >=latex] (\from) -- node[weight] {$\weight$} (\to);

	\end{tikzpicture}
	\caption{Graphical representation of an XSD containing references to global elements.}%
	\label{fig:xsd-shared-components}
\end{figure}
In the diagram of Figure \ref{fig:xsd-shared-components},  the XSD is modeled as a graph with a loop.
}
$\punto$
\end{example}

XSD offers further opportunities to reuse the definition of a schema (or of a part of it) in other schemas. In
particular, an XSD $S_1$ (called {\em source schema}) can be chained to another XSD $S_2$ (called {\em target schema})
so that the elements, the attributes and the complex types defined in $S_1$ can be reused in $S_2$. There are two ways
to reuse the source schema in the target one; they are based on the {\tt import} and {\tt include} directives. The
former assumes that the namespace of the source schema differs from that of the target one. The latter assumes that the
source and the target schemas share the same namespace.

We can illustrate these concepts on the University Library example.

\begin{example}
\label{ex:include}
{\em Suppose that a file {\tt Book.xsd}, containing the specification of the element {\tt Book} described in Example
\ref{ex:shared}, was generated. We can reuse the definition of a {\tt Book} to generate a catalogue of books. In Figure
\ref{fig:includeschema} we report the XSD {\tt BookCatalogue.xsd}, describing the structure of a catalogue of books. If
we assume that {\tt Book.xsd} and {\tt BookCatalogue.xsd} share the same namespace, we can include {\tt Book.xsd} into
{\tt BookCata\-logue.xsd}, thus avoiding to redefine the element {\tt Book}. Of course, {\tt Book.xsd} can be used to
define further schem\-as (e.g., the books purchased by the Library or borrowed by students).

\begin{figure}[t]
\centering
{\scriptsize
\begin{lstlisting}
<?xml version="1.0"?>
<xsd:schema xmlns:xsd="http://www.w3.org/2001/XMLSchema"
            targetNamespace="http://www.library.org"
            xmlns="http://www.library.org"
            elementFormDefault="qualified">
    <xsd:include schemaLocation="Book.xsd"/>
    <xsd:element name="Library">
        <xsd:complexType>
             <xsd:sequence>
                 <xsd:element name="BookCatalogue">
                     <xsd:complexType>
                         <xsd:sequence>
                             <xsd:element ref="Book"
                                   maxOccurs="unbounded"/>
                         </xsd:sequence>
                     </xsd:complexType>
                 </xsd:element>
            </xsd:sequence>
        </xsd:complexType>
    </xsd:element>
</xsd:schema>
\end{lstlisting}
}
\caption{The XSD {\tt BookCatalogue.xsd}, describing a catalogue of books.}
\label{fig:includeschema}
\end{figure}
}$\punto$
\end{example}

A designer may opt to define multiple XSDs and to assembly them (by means of the {\tt import}/{\tt include} directives)
for obtaining a new XSD $S$. In this way, the namespace of $S$ is {\em distributed across} multiple namespaces, each
corresponding to an XSD used in the definition of $S$.

The reuse of a schema, a sub-schema or a single element can reveal semantic relationships between the involved parties.
For instance, if two elements refer the same shared component, we can assume that a semantic relationship exists
between them (think of elements {\tt Book} and {\tt Article} discussed in Example \ref{ex:shared}). Ideally, an XML
Matcher should be able to identify semantic matchings independently of the design style adopted to construct the XSD.
It clearly emerges, from the previous reasoning, that discovering semantic matchings in presence of shared components
and/or distributed namespaces is extremely challenging and difficult.
%
%
%

\section{A template to classify XML Matchers}
\label{sec:conceptual}

In this section we introduce a template to classify {\em XML Matchers}. This task is intricate because of
two main reasons.

The first one is that there is not a general agreement about the definition of XML Matcher and its goals. In fact, for
some authors \cite{Ursino-IS4}, an XML Matcher can be conceived as a function which receives two schema elements
belonging to two different DTDs/XSDs and returns {\tt true} if a semantic matching exists between them, {\tt false}
otherwise. In compliance with Definition \ref{def:matching}, the output of an XML Matcher could be a real number in
$[0,1]$, stating the semantic similarity of the corresponding schema elements so that the higher this number the more
likely they form a matching. Other authors \cite{WoMlDo10,SmKeJo05}, instead, conceive XML Matchers as tools to compute
the {\em similarity degree} of two DTDs/XSDs; this degree is a numerical value stating to what extent the semantics
captured by the two schemas overlap. A nice application of such a definition can be found in the {\em BellFlower}
system \cite{SmKeJo05}. Here, the authors introduce the concept of {\em personal schema querying}, i.e. they assume
that a repository of XML documents is available on the Web and consider a scenario in which a user wants to query it
but is unfamiliar with the structure of available documents and, therefore, is not able to formulate precise queries.
This user first can provide her own virtual view of unknown data; this view is encoded as an XSD ({\em personal
schema}). Then, an XML Matcher ranks the XSDs in the repository according to their similarity with the personal schema.

The second reason is that researchers, working independently of each other, adopted different methodologies and tools
to find semantic matchings between DTDs/XSDs. At a first glance these approaches appear unrelated to each other
because, for instance, some of them heavily rely on Machine Learning techniques, whereas others are based on graph
theory. As a consequence, a time-based framework to classify existing approaches in such a way that one approach
attempts to solve the disadvantages of the previous ones is hard to define and, perhaps, not effective in describing
how research on XML Matchers has evolved over time.

In order to solve the challenges outlined above, we introduce an {\em abstract model} which plays the role of {\em
template} whose structure is applicable to all existing XML Matchers. Our template consists of several {\em
components}; for instance, a component could specify the strategy adopted for representing input DTDs/XSDs, whereas
another could describe the strategy for deciding whether two schema elements are to be regarded as similar or not. Once
these components have been defined, we can represent each existing approach as a set of them. This way, unrelated
approaches differ for the strategies implemented by them in one or more components. Such a template acts as a {\em
universal container} which is generally enough to describe the main features of most of the existing XML Matchers. In
the following we will call it as {\em XML Matcher Template}.

To formally introduce it, we need to provide some preliminary definitions. In particular, we call {\em Input Format Domain}
(and we indicate it as $\mathtt{IFD}$) the set of the elements provided in input to the Matcher. So, for instance, in
some approaches the $\mathtt{IFD}$ coincides with a set of XML Schemas \cite{Ursino-IS4,MaBeRa01}, in other ones it
coincides with a set of DTDs \cite{Lee*02} and, finally, in other ones it coincides with a set of XML documents plus
the DTDs/XSDs associated with them \cite{DoDoHa01}.

As a further, preliminary concept, we observe that XML Matchers often perform a pre-processing step on input DTDs/XSDs;
for instance, they convert DTDs/XSDs into trees \cite{WoMlDo10} or graphs \cite{Ursino-IS4,MaBeRa01}. Such a
pre-processing step can be seen as the output of a {\em function} from $\mathtt{IFD}$ to an {\em Internal Representation Domain}
(that we indicate as $\mathtt{IRD}$). On the basis of the definitions above, we are now able to introduce the concept of
XML Matcher Template.

\begin{definition} ({\em XML Matcher Template}).
\label{def:xmlmatcher}
{\em An {\em XML Match\-er Template} $\mathcal{T}$ is a tuple $\mathcal{T} = \langle \mathtt{IFD}, \mathtt{IRD},$
$\mathcal{M}, k, \overrightarrow{\sigma}, \lambda, \phi \rangle$ where:

\begin{enumerate}

\item $\mathtt{IFD}$ and $\mathtt{IRD}$ are the Input Format and the Internal Representation domains, respectively.

\item $\mathcal{M}: \mathtt{IFD} \rightarrow \mathtt{IRD}$ ({\em pre-processing function}) is a function which receives an
    element of $\mathtt{IFD}$ and returns an element of $\mathtt{IRD}$.
    For each $S \in \mathtt{IFD}$, we denote as $\mathcal{M}(S)$ the corresponding element in $\mathtt{IRD}$.

\item $k$ is an integer greater than or equal to 1.

\item $\overrightarrow{\sigma} = \left[\sigma_1, \ldots, \sigma_k \right]$ is a group of $k$ {\em functions}. For
    each $i = 1 \ldots k$, and for each $S \in \mathtt{IFD}$, $\sigma_i: \mathcal{M}(S) \times \mathcal{M}(S) \rightarrow \left[0,1\right]$  is called {\em similarity function}.

\item $\lambda : \left[0,1\right]^{k} \rightarrow \left[0,1\right]$ is an {\em aggregating function};
    $\left[0,1\right]^{k}$ denotes the hypercube in $\mathbb R^k$, i.e. a $k$-th dimensional array whose components
    range from 0 to 1.

\item $\phi : \mathtt{IFD} \times \mathtt{IFD} \rightarrow \R^+$ is a {\em schema similarity function}; $\R^{+}$ is
    the set of non-negative real numbers.

\end{enumerate}
}
$\punto$
\end{definition}

The similarity function $\sigma(\cdot,\cdot)$ receives two schema elements and returns a real number in $[0,1]$ ({\em
similarity score}). The output of $\sigma(\cdot,\cdot)$ plays the role of the confidence measure $\overline{c}$
introduced in Definition \ref{def:matching}; $\sigma(\cdot,\cdot)$ is characterized by the following further
constraints:

\begin{enumerate}
\item For each $u$, $\sigma(u,u) = 1$, i.e. each element is similar to itself with the highest similarity score.

\item For each pair of schema elements $\langle u, v\rangle $ such that $u$ (resp., $v$) belongs to a DTD/XSD $S_1$
    (resp., $S_2$), $\sigma(u,v) = \sigma(v,u)$, i.e. $\sigma(\cdot, \cdot)$ is {\em symmetric} with respect to its
    arguments.

\end{enumerate}

In the following, when it does not generate confusion, we will use the symbols $u$ and $v$ to denote two schema
elements of two different DTDs/XSDs. In case we need to handle more elements in $S_1$ (resp., $S_2$), we will denote
them as $u_1, u_2, \ldots, u_n$ (resp., $v_1, v_2, \ldots, v_n$).

The {\em aggregating function} $\lambda$ is necessary because, in most cases, one can define different similarity
functions to compute the similarity score of two schema elements. This depends on the fact that, in general, there is
no similarity function which always returns better results than the other ones and, therefore, it can be convenient to
use more similarity functions \cite{AlNaSa10}. $\lambda$ has the purpose of aggregating the results generated by these
functions to yield a {\em global similarity score}.

Finally, the function $\phi$ receives a pair of DTDs/XSDs and returns a non-negative real number. The goal of $\phi$ is
to compute to what extent two DTDs/XSDs describe the same piece of reality. It is usually based on $\sigma$ and
$\lambda$, i.e., the similarity degree of two schemas is computed by assembling the similarity scores of
their elements. A nice application of $\phi$ is in the field of XML Schema Clustering (see Section
\ref{sub:schemaclustering}).

In the next subsections we will comparatively discuss the features of some popular XML Matchers in terms of the XML
Matcher Template. In particular, we will discuss the XML Matchers proposed in
\cite{DoDoHa01,MaBeRa01,Aumuller*05,Ursino-IS4,Lee*02,WoMlDo10,AlNaSa10,YiHuCh05,Kim*11,Jeong*08}.

Observe that both {\tt LSD} \cite{DoDoHa01}, {\em Cupid} \cite{MaBeRa01} and {\em COMA++} \cite{Aumuller*05} have been conceived to handle generic
schemas, but they are included in our review because they provide some specific modules for the management of
DTDs/XSDs.

\subsection{The pre-processing function $\mathcal{M}$}
\label{sub:mapping}

In this section we consider how the mapping function $\mathcal{M}$ is implemented in the XML Matchers into examination.

In {\tt LSD} \cite{DoDoHa01}, the $\mathtt{IFD}$ coincides with a set of XML documents and the DTDs associated with
them. As observed in Section \ref{sec:matching-xml}, {\tt LSD}, in its basic version, does not consider the
hierarchical structure of a document. Indeed, an XML document is mapped onto an array whose elements are called {\em
tokens}. Therefore, the $\mathtt{IRD}$ is the space of $n_t$-th dimensional arrays, being $n_t$ the number of tokens in
a document. A variant of {\tt LSD} has been designed to deal with hierarchical data models and, in particular, with
DTDs. In that case, two kinds of token, namely {\em node tokens} and {\em edge tokens}, are considered. A node token is
an element of the DTD/XSD or an instance of that element, whereas an edge token specifies a hierarchical relationship
between schema elements.

Apart from {\tt LSD}, all the other discussed approaches map DTDs/XSDs onto trees or graphs.

\begin{table*}
\centering

{\scriptsize
\begin{tabular}{||c|c|c|c|c||}
\hline  \hline
{\em Approach}                      & $\mathtt{IFD}$    & $\mathtt{IRD}$ & {\em Element} & {\em Relationships in }\\
                                    &                   &                & {\em Labels}  & $\mathtt{IFD}$ \\
\hline \hline
{\tt LSD}                     & XML Documents +      & An array of tokens /     & Name & -\\
                              & DTDs             & A pair of arrays of tokens &      &\\

\hline
Cupid                         & DTDs/XSDs                & Rooted Graphs   & Name & Aggregation, \\
                              &                          & and Trees       &      & Generalization and\\
                              &                          &                 &      & Specialization\\
\hline

COMA++                        & DTDs/XSDs                & Directed   & Name & Aggregation, \\
                              &                          & Acyclic    &      & Generalization and\\
                              &                          & Graphs     &      & Specialization\\
\hline

XIKE                          & XSDs               & Directed        & Name & Element/Sub-element\\
                              &                          & Graphs          &      & Element/Attributes\\
\hline
XClust                        & DTDs                 & Trees            & Name & Element/Sub-Element\\
\hline
Approach of \cite{WoMlDo10}   & DTDs/XSDs                 & Trees            & Name & Element/Sub-Element\\
\hline
Approach of \cite{AlNaSa10}   & XSDs               & Trees +           &  Name, Data Type  & Element/Sub-Element\\
                              &                          & Prufer Sequences  &  and Cardinality  & \\

\hline
Approach of \cite{YiHuCh05}   & XSDs                & Directed         & Name, Attributes, Parents, & Element/Sub-Element\\
                              &                           & Graphs           & Children and Brothers        & \\

\hline
Approach of \cite{Kim*11}     & DTDs                  & Trees            & Name                     & Element/Sub-Element \\
\hline
Approach of \cite{Jeong*08}   & DTDs                  & Trees            & Name                     & Element/Sub-Element \\
\hline \hline
\end{tabular}

}

\caption{A classification of the XML Matchers into evaluation in terms of $\mathtt{IFD}$, $\mathtt{IRD}$, labels attached
with schema elements and relationships among elements in $\mathtt{IRD}$.} \label{tab:mapping}

\end{table*}

In XClust \cite{Lee*02} and similar methods \cite{WoMlDo10}, the $\mathtt{IFD}$ is a set of DTDs. Each schema element in
the original DTD is mapped onto a tree node, whereas an edge models a relationship between an element and one of its
sub-elements. Due to the presence of repeated and shared elements, DTDs may assume the form of graphs with loops. Some
rules are required to break loops.

Algergawy et al. \cite{AlNaSa10} focus on XSDs which are mapped onto trees called {\em Schema Trees}. Each element of
an XSD is mapped onto a (unique) node in the Schema Tree; edges are used to model parent relationships. Each node is
provided with a {\em label} reporting its name, its data type and its cardinality constraints. Also in \cite{Jeong*08}
and \cite{Kim*11}, DTDs are mapped onto trees such that each node is labeled with the element name and each edge
encodes an element/sub-element relationship.

Other approaches map the DTDs/XSDs provided in input onto graphs. In Cupid \cite{MaBeRa01}, the $\mathtt{IRD}$ is a {\em
rooted graph} such that each node uniquely corresponds to a schema element. Elements are tied by different types of
edges encoding relationships, like generalization, specialization and aggregation, discussed in Section
\ref{sec:matching-xml}. Finally, a set of procedures for converting graphs into trees is defined.

As for COMA++, it converts DTDs/XSDs onto directed acyclic graphs; each element in the original schema is represented by means of the path joining the schema root with the element itself.

In XIKE \cite{Ursino-IS4}, the $\mathtt{IFD}$ is a set of XSDs whereas the $\mathtt{IRD}$ is a set of directed and
labeled graphs called {\em XS-Graphs}. Each node of an XS-Graph is uniquely associated with an element of an XSD,
whereas an edge linking two nodes specifies, for instance, an element/sub-element relationship or an element/attribute
relationship.

Yi et al. \cite{YiHuCh05} propose to model an XSD as a {\em semantic network}, i.e. as a graph with labeled nodes. As
in the previous approaches, each element in the XSD is uniquely associated with a node in the graph. Each node $v$ is
provided with the following functions: {\em (i) name(v)} - $n(v)$, returning the name of $v$; {\em (ii) attributes(v)}
- $a(v)$, returning the attributes of $v$; {\em (iii) parent(v)} - $p(v)$, returning the parent element of $v$; {\em
(iv) children(v)} - $c(v)$, returning the set of child elements of $v$; {\em (v) brother(v)} - $b(v)$, returning the
set of brother elements of $v$. Two elements are said {\em brothers} if they share the same parent.

In Table \ref{tab:mapping} we report a classification of the discussed approaches in terms of $\mathtt{IFD}$ and $\mathtt{IRD}$, of the label associated with each schema element in $\mathtt{IRD}$ and, finally, of the relationships among
elements encoded in $\mathtt{IRD}$.

\subsection{The similarity function $\sigma$}
\label{sub:similarity}

The similarity function $\sigma$ plays a key role in the whole matching process; therefore, we will describe its
features in details.

According to Definition \ref{def:xmlmatcher}, the function $\sigma$ operates on elements of $\mathtt{IRD}$; observe
that different functions $\sigma$ can be defined. Existing definitions of $\sigma$ fall within the following classes:

\begin{itemize}

\item A first group considers information about elements (like data types or name elements). We call them as {\em
    element-level Matchers}.

\item A second group analyzes the structural information included in DTDs/XSDs. These Matchers are based on the
    concept of {\em context} of an element; for instance, if an XSD is modeled as a tree, the context of an element
    consists of its descendants and ancestors. In this contexts it is assumed that there exists a semantic matching
    between two nodes if there exists a matching between their contexts. Approaches belonging to this category are
    called {\em context-level Matchers}.

\end{itemize}

In the following subsections we will present in detail these two classes of XML Matchers.
We pay a special attention on COMA++: a key feature of this system is, in fact, its ability of implementing several similarity functions $\sigma$.
These functions can be applied in parallel (and in an independent fashion) under user control.
The first version of COMA++ supported more than 10 version of the $\sigma$ function; some of these implementations were merely linguistic (i.e., considered only the names of elements to match) while other were structural (i.e., considered the context of two elements).
In principle COMA++ can be extended in such a way as to implement any of the approaches we will present in the next sections and, therefore, we do not report our discussion about COMA++.

\subsubsection{Element-level Matchers}
\label{subsub:Element-Level-Matching}

Approaches belonging to this category use information like the names of the elements, their data types and the
constraints on their cardinalities to compute matchings; each of these features provides a basic similarity metric.

A first category of similarity metrics can be classified as {\em Name Similarity}. The corresponding metrics are based
on the following intuition: ``{\em The more similar the names of two XML elements, the higher the similarity level of
these last ones.}''.

Therefore, to assess the similarity of two schema elements, it is necessary to compare the strings representing their
names. To this purpose, various methods, like {\em prefix}, {\em suffix}, {\em edit distance}, {\em Jaro distance} and
{\em n-grams} \cite{AlScSa09,GiYa04}, have been proposed in the literature. In the following we will provide an
overview of them:

\begin{itemize}

	\item {\em Prefix}. Prefix receives two strings and checks whether the first one starts with the second one.
Prefix is efficient in recognizing {\em acronyms} (e.g., it recognizes that terms like ``int'' or ``integer'' form
a semantic matching). Prefix strategy is implemented in Cupid.

	\item {\em Suffix}. Suffix receives two strings and checks whether the first one ends with the second one
(e.g., ``phone'' and ``telephone''). Both the systems presented in \cite{DoRa02,MeGaRa02} and Cupid exploit this
metrics.

	\item {\em Edit distance}. Edit or Levenshtein distance \cite{Levenshtein66} receives two strings and computes
the shortest sequence of operations (like insertion/deletion of characters) capable of transforming the former into
the latter. Edit distance has been used in \cite{DoRa02}. In the context of DTDs, it was adopted in the approach of
\cite{WoMlDo10}. This distance is useful when abbreviations, instead of complete names, are used. The resulting
similarity metric is called {\em Syntactic Similarity} and denoted as $SynSim(\cdot,\cdot)$.

	\item {\em Jaro distance} \cite{Jaro95}. Given two strings $s_1$ and $s_2$, their {\em Jaro distance} $d_J$is
defined as follows:

$$
d_J = \frac{1}{3} \cdot \left( \frac{m}{|s_1|} + \frac{m}{|s_2|} + \frac{m-t}{m} \right)
$$

\noindent being $m$ the number of matching characters between $s_1$ and $s_2$, $t$ the number of {\em character
transpositions}, i.e. the number of characters matching in $s_1$ and $s_2$, but in a different order, and $|s_1|$
(resp., $|s_2|$) the length of $s_1$ (resp., $s_2$). Jaro distance is adopted in \cite{QuKeLi07}.

	\item {\em N-gram}. An $n$-gram is a sequence of $n$ consecutive characters in a string. For instance, given
the string ``house'', the $n$-grams of length 3 associated with it (often known as {\em trigram}) are ``hou'',
``ous'' and ``use''. In approaches relying on $n$-grams, the distance between two strings is computed by comparing
the number of $n$-grams shared by them. In the context of Schema Matching, examples of approaches exploiting
$n$-grams are reported in \cite{DoRa02,GiShYa05}.

\end{itemize}

An experimental comparison of different metrics for string matching is reported in \cite{CoRaFi03}.

A second category of metrics can be classified as {\em Language-Based Similarity}. They use Natural Language Processing
tools, like tokenization and elimination, already introduced in Section \ref{sub:basic}. In the context of DTDs/XSDs,
Language-Based Similarity metrics have been applied, in their basic form, in Cupid and in the approach of
\cite{Jeong*08}. Some authors suggest that, in order to compute the similarity of two XML elements, it is necessary to
consider not only the names of the elements themselves but also the names of their children or their parents
\cite{YiHuCh05}. For instance, let us consider the contribution to the similarity score computation provided by the
children elements. Given two schema elements $u$ and $v$, let $c(u)$ and $c(v)$ be the children of $u$ and $v$,
respectively. The similarity score $\sigma_{c}(u,v)$ is defined as follows \cite{YiHuCh05}:

\begin{equation}
\sigma_{c}(u,v) = \frac{N_{\cap}}{N_{\cap} + \alpha_{uv}N_{u-v} + \left(1- \alpha_{uv}\right)N_{v-u}} \label{eqn:similaritychildren}
\end{equation}

\noindent where $N_{\cap}$ is the number of children of $u$ which are also children of $v$, and $N_{u-v}$ (resp.,
$N_{v-u}$) is the number of children of $u$ (resp., $v$) which are not children of $v$ (resp., $u$). The coefficient
$\alpha_{uv}$ is a weight ranging in $[0,1]$.

Equation \ref{eqn:similaritychildren} can be generalized to the other element features, like attributes. The reason
underlying this kind of similarity is as follows: The higher the number of features shared by $u$ and $v$, the higher
their similarity degree. By contrast, a high number of non-shared features of $u$ and $v$ induces a strong penalization
on their similarity degree. Since each element has exactly one parent, Equation \ref{eqn:similaritychildren}, in case
of parent function $p(v)$, becomes as follows:

\begin{equation}
\label{eqn:similarityparent}
\sigma_{p}(u,v)=
\begin{cases} 1 & \text{if $p(u) = p(v)$,}
\\
0 &\text{if $p(u) \neq p(v)$.}
\end{cases}
\end{equation}

\noindent being $p(u)$ (resp., $p(v)$) the unique parent of $u$ (resp., $v$). Most of the Language-Based Similarity
metrics rely on the use of {\em external sources}, such as dictionaries or thesauri, like WordNet. WordNet has been
exploited, in the context of XSDs, in XIKE \cite{Ursino-IDEAS2004,Ursino-Informatica1,Ursino-IS4} and other methods
\cite{AlNaSa10}, whereas it was adopted, in the context of DTDs, in XClust \cite{Lee*02} and other approaches
\cite{WoMlDo10}. In \cite{WoMlDo10}, the authors define a similarity measure obtained by querying WordNet. This
similarity is called as {\em Semantic Similarity}, and is denoted as $SemSim(\cdot,\cdot)$.

In some cases, like in {\tt LSD}, an external source is constructed by taking the matching manually provided by users
into account. In particular, in {\tt LSD}, the problem of finding semantic matchings is formulated as a {\em
classification problem}: Given a set of pre-defined classes $c_1, \ldots, c_m$, two schema elements form a semantic
matching if they share the same class. To perform classification, {\tt LSD} provides two main alternatives: The former
is based on {\sf Whirl} \cite{CoHi98}, a $k$-nearest-neighbor text classifier, whereas the latter exploits the {\em
Naive Bayes classifier}. To train these classifiers, {\tt LSD} looks at the instances of two elements and applies the
TF-IDF metric on them. In other words, in {\tt LSD}, the similarity score of two schema elements depends on the
similarity of their instances. A further classifier, capable of considering the hierarchical structure of XML documents
and DTDs, is included in {\tt LSD}; it will be discussed in Section \ref{sub:Context-Level-Matching}.

A third category of metrics is given by {\em Data Type Similarity}. These metrics depend on type constraints (see
Section \ref{sub:constraint-data-type}). Data Type Similarities are considered in {\em Cupid}, XIKE and in other
approaches \cite{AlNaSa10}.

Finally, a fourth category of metrics is based on the concept of {\em Cardinality Constraint Similarity}. A discussion
about cardinality compatibility and its usage in the matching of DTDs/XSDs has been performed in Section
\ref{sub:cardinalities}. As for DTDs, cardinality constraints are considered in XClust and in the approach of
\cite{WoMlDo10}: In this latter approach, the function evaluating the similarity degree of two elements on the basis of
the compatibility of their cardinalities is called $CardSim(\cdot,\cdot)$. The usage of cardinality constraints in XSDs
is considered in \cite{AlNaSa10} and in XIKE.

In Table \ref{tab:elementlevel} we classify the approaches into evaluation on the basis of the element-level Matchers
they adopt.

\begin{table*}
\centering

{\scriptsize
\begin{tabular}{||c|c|c|c|c||}
\hline  \hline
{\em Approach}                      & {\em Language-Similarity}  & {\em Data Type} & {\em Cardinality Constraint}\\
\hline \hline
{\tt LSD}                     & TF/IDF & No & No\\
\hline
Cupid                         & Tokenization/Elimination & Yes & No \\
\hline
XIKE                          & WordNet & Yes & Yes\\
\hline
XClust                        & WordNet & Yes & Yes\\
\hline
Approach of \cite{WoMlDo10}   & Edit Distance & No & Yes \\

\hline
Approach of \cite{AlNaSa10}   & Tokenization/Elimination & Yes & Yes\\
\hline
Approach of \cite{YiHuCh05}   & Name comparison (children, &  No & No \\
                              & brother, parent)           &  & \\

\hline
Approach of \cite{Kim*11}     & WordNet & No  &  No\\
\hline
Approach of \cite{Jeong*08}   & Tokenization/Elimination & No & No \\
\hline \hline

\end{tabular}
} \caption{A classification of the XML Matchers into evaluation in terms of the element-level Matchers they implement.}
\label{tab:elementlevel}
\end{table*}

\subsubsection{Context-level Matchers}
\label{sub:Context-Level-Matching}

Context-level Matchers are based on the fact that a DTD/XSD is usually represented as a tree $T$ or a graph $G$, and
each of its elements corresponds to a node in $T$ or $G$. The {\em context} of a node $n$ in $T$ (resp., $G$) is
represented by all the nodes of $T$ (resp., $G$) ``close'' to $n$, according to some closeness definition.

The context of $n$ is exploited to better interpret the meaning of the element represented by $n$. The intuition behind
context-level Matchers is as follows: ``{\em If two elements are similar, their contexts should also be somehow
similar}''. In the field of Schema Matching such an idea was first introduced in \cite{FaKrNe91}; furthermore, it is
popular in other fields of Computer Science, like {\em Data Mining} \cite{JeWi02} and {\em Social Network Analysis}
\cite{LiKl03}.

Context-level Matchers can be classified into two categories, namely: {\em (i)} {\em tree-based approaches}, if the
DTD/XSD is modeled as a tree; {\em (ii)} {\em graph-based approaches}, if the DTD/XSD is modeled as a graph.

\paragraph {\sf Tree-based approaches.} In tree-based approaches, the context of a node usually coincides with the set
of its ancestors and descendants. This implies that similarities are computed only for {\em non-leaf nodes} (in fact, a
leaf node has only one parent and no descendants). Tree-based approaches can be classified as follows \cite{AlNaSa10}:

\begin{itemize}

	\item {\em Child similarity}. These approaches consider only the children of the nodes into examination. Two
non-leaf nodes are classified as {\em similar} if their sets of children present a high matching degree. The sets
of children of two nodes could be compared by applying the Jaccard coefficient or other analogous methods. An
implementation of this strategy can be found in Cupid.

	\item {\em Leaf similarity}. In these approaches, any non-leaf node $n$ is considered as the root of a {\em
subtree}. The similarity degree between two non-leaf nodes $n_1$ and $n_2$ is computed by comparing the sets of the
leaf nodes belonging to the subtrees having $n_1$ and $n_2$ as root. As an example, in \cite{AlNaSa10,AlScSa09},
first leaf nodes are mapped onto {\em arrays} of real numbers and, then, the similarity of these arrays is computed
by applying the cosine similarity measure.

	\item {\em Sibling similarity}. The siblings of a node $n$ are the nodes placed at the same level as $n$ in the
tree \cite{AlNaSa10}. Given two nodes $n_1$ and $n_2$, first these approaches construct the sets of their siblings.
After that, they compute the similarity of each pair of siblings and, then, select the pairs with the highest
similarity scores ({\em best matching pairs}). Finally, they compute the similarity degree of $n_1$ and $n_2$ by
averaging the similarity degrees of the best matching pairs.

	\item {\em Ancestor similarity}. In these approaches, the ancestors of a node are considered. In particular,
given a node $n$, let $\mathtt{path}(n)$ be the path joining the root with $n$. The set of the nodes in
$\mathtt{path}(n)$ is called the {\em ancestor context} of $n$. These approaches, in order to compute the
similarity degree of two nodes $n_1$ and $n_2$, compare the corresponding ancestor contexts. Some authors
\cite{AlScSa09} propose to use refined techniques (like {\em Prufer sequences} \cite{Prufer18}) to efficiently
encode these paths.

\end{itemize}

As discussed in Section \ref{sub:mapping}, {\tt LSD} encodes XML documents on the basis of their hierarchical
structure. Once each XML document has been mapped onto a pair of vectors, a classifier is applied to find matchings.
Cupid implements an approach which combines both {\em Leaf Similarity} and {\em Ancestor Similarity}. In particular,
two non-leaf elements are classified as similar if they are linguistically similar and the sub-trees having the two
elements as root are similar. In an analogous fashion, two non-leaf schema elements are structurally similar if their
leaves are highly similar. This leads to a {\em mutually recursive} definition of similarity: Two elements are similar
if their leaves are similar, and the similarity of two leaves depends on the similarity of their ancestors.

An idea similar to that proposed in Cupid is explored in XClust, where the context of an element consists of both its
ancestors (assuming that this element is not the root) and its descendants. The descendants of an element consist of
its attributes, its sub-elements, the elements linked to it by means of {\tt IDREF(S)} attributes, as well as the
leaves of the subtree rooted at the element itself.

In \cite{Kim*11}, the problem of finding semantic matchings is formulated as the problem of finding a matching between
two trees. This approach first reduces the tree-to-tree matching problem to a path-to-path matching problem which, in
its turn, is formulated as a node-to-node matching problem. Finally, this last problem is solved as a word-to-word
matching problem. Each of these problems can be suitably formulated as a maximum weight matching problem on a bipartite
graph with different constraints. To deal with computational complexity, integer programming and dynamic programming
techniques are applied.

\paragraph{\sf Graph-based approaches.} These approaches consider input schemas as
graphs (often weighted and labeled).
Relationships between schema elements are modeled by structural properties (like edges or paths) of a graph.

In XIKE, XSDs are converted into labeled graphs called XS-Graphs. A distance between two schema elements
can be defined on the basis of the length of the shortest path linking the corresponding nodes in the associated
XS-Graph. In XIKE, the {\em q-neighborhood of a node $n$} consists of the set of the nodes whose distance from $n$ is
less than a pre-fixed threshold $q$.

XIKE takes an integer $q$ (called {\em severity level}) as well as a pair of schema elements $u$ and $v$ as input. Let
$n_u$ (resp., $n_v$) be the node corresponding to $u$ (resp., $v$) in the XS-Graph associated with the XSD containing
$u$ (resp., $v$). A bipartite graph $BG = \langle N_u \cup N_v, E \rangle$ is built such that: {\em (i)} $N_u$ is the
$q$-neighborhood of $n_u$, {\em (ii)} $N_v$ is the $q$-neighborhood of $n_v$, and {\em (iii)} an edge $e \in E$ links a
node of $N_u$ with a node of $N_v$ if there exists a {\em syntactic matching} (revealed by querying WordNet) between
them. After that, a {\em maximum weight matching problem} on $BG$ is solved and, if the value of the corresponding
objective function is higher than a threshold then $u$ and $v$ are considered semantically similar. XIKE is {\em
parametric} against the severity level $q$. It is {\em scalable} because the size of $BG$ is usually small.

In \cite{YiHuCh05}, a relaxation labeling algorithm is proposed to find semantic matchings in XSDs. Relaxation labeling
is an efficient technique to solve the problem of assigning labels to the nodes of a graph so as to satisfy a set of
constraints. Two schema elements are considered semantically related if they share the same label. The approach of
\cite{YiHuCh05} considers {\em structural} as well as {\em semantic} constraints.

\subsection{The aggregating function $\lambda$}
\label{sub:lambda}

The similarity functions defined before exploit only a specific feature of schema elements (e.g., name or context).
Therefore, the usage of a single strategy cannot be sufficient because a similarity function can produce accurate
results in a particular application domain, but it can work poorly in other ones \cite{AlNaSa10,DoDoHa01,MaBeRa01}. For
this reason, several authors suggest to simultaneously use several similarity functions. In this way, for a fixed pair
of schema elements, multiple similarity scores (called {\em partial similarity scores}) are available. These scores
are, then, {\em combined} (or {\em aggregated}) into a {\em global similarity score}
\cite{AlNaSa10,DoDoHa01,Jeong*08,MaBeRa01}.

On one hand, aggregating partial similarity scores provides a high level of adaptability because potential errors of a
similarity function are compensated by the other ones at disposal of the Matcher. This allows the improvement of the
overall accuracy. On the other hand, combining (or aggregating) partial similarity scores is not trivial, and several
strategies have been proposed so far \cite{AlNaSa10}.

By inspiring ourselves with the ideas proposed in \cite{AlNaSa10}, we suggest to classify aggregation strategies into
{\em homogeneous}, if the similarity functions are of the same type (e.g., all of them operate at the element-level),
and {\em heterogenous}, otherwise (e.g, a similarity function works at the element-level and the other one operates at
the context-level).

In some cases, the computation of $\lambda$ can be seen as an {\em iterative process} \cite{Jeong*08}. A human expert
checks the results generated by the Matcher and, if they are not in line with her expectations, can adjust the
thresholds used in the similarity functions, can decide to consider further similarity functions (in addition to the
already used ones), can remove some similarity functions from the pool of the considered ones, or, eventually, can
modify the aggregation strategy.

To formally introduce the concept of aggregating function, we assume that $k \geq 1$ similarity functions
$\sigma_i(\cdot,\cdot)$ are available. We define the hypercube $\mathcal{C}$ in $\R^k$ (and denote it as $[0,1]^k$) as
a vector of $k$ components, each belonging to the real interval $[0,1]$. A vector $\overrightarrow{x} = \{x_1, \ldots,
x_k \}$ {\em belongs to} $\mathcal{C}$ (and we will write $\overrightarrow{x} \in [0,1]^k$) if, for each $i = 1 \ldots
k$, $0 \leq x_i \leq 1$. For each pair of vectors $\overrightarrow{x} \in [0,1]^k$ and $\overrightarrow{y} \in
[0,1]^k$, we say that $\overrightarrow{x} \preceq \overrightarrow{y}$ (i.e., $\overrightarrow{x}$ precedes
$\overrightarrow{y}$) if $x_i \leq y_i$ for each $i = 1 \ldots k$. This way, a precedence relationship $\preceq$ is
defined.

We are now able to define an {\em aggregating function} as in \cite{BePrCa07}:

\begin{definition}({\em Aggregating Function})
\label{def:aggregating}
{\em An {\em aggregating function} $\lambda: \R^k \rightarrow [0,1]$ is a function that satisfies the following two
properties:

\begin{enumerate}

\item $\lambda(\overrightarrow{0}) = 0$ and $\lambda(\overrightarrow{1}) = 1$, being $\overrightarrow{0}$ (resp.,
    $\overrightarrow{1}$) the $k$-th dimensional vector whose components are all equal to 0 (resp., 1).

\item For each $\overrightarrow{x} \in [0,1]^k$ and $\overrightarrow{y} \in [0,1]^k$, if $\overrightarrow{x}
    \preceq \overrightarrow{y}$, then $\lambda(\overrightarrow{x}) \leq \lambda(\overrightarrow{y})$.

\end{enumerate}
}$\punto$.
\end{definition}

The condition $\lambda(\overrightarrow{0}) = 0$ means that if two elements are recognized as totally dissimilar by all
the available similarity functions, then the global similarity score must be 0. In an analogous fashion, if all the
available similarity functions agree on the fact that the similarity degree of two schema elements achieves its highest
value then the aggregating function must return 1.

Condition 2 states that $\lambda$ is {\em monotonic non-decreas\-ing}. The monotonicity of $\lambda$ must be
interpreted as follows: Let us consider two pairs of elements $\langle u_1, v_1 \rangle$ and $\langle u_2, v_2 \rangle$
and suppose that $\sigma_i(u_1,v_1) \leq \sigma_i(u_2,v_2)$ for all $i = 1 \ldots k$, i.e. that all similarity
functions agree on the fact that the similarity score of the pair $\langle u_1, v_1 \rangle$ is less than or equal to
the similarity score of the pair $\langle u_2, v_2 \rangle$. In such a case, the definition of $\lambda$ must guarantee
that $\lambda(u_1,v_1)$ is less than or equal to $\lambda(u_2,v_2)$.

The simplest way of aggregating partial similarity scores is to compute their weighted sum; in this case $\lambda$ can
be defined as follows:

\begin{equation}
\label{eqn:weightedsum}
\lambda(u,v)= \sum_{i = 1}^k w_i \sigma_i(u,v)
\end{equation}

Here, a weight $w_i$ can be interpreted as the {\em confidence} on the correctness of the results produced by
$\sigma_i(\cdot,\cdot)$.
Some systems, like COMA++ \cite{Aumuller*05}, consider also the configuration in which weights are all equal (i.e., $w_i = w^{*} \forall i$) and in such a case $\lambda(u,v)$ is equal to the average of the similarity scores generated by each $\sigma_i$.

In some cases, the weights $w_1 \dots w_k$ form a {\em convex combination}, i.e. $\sum_{i=1}^k w_i = 1$. This is the
case of the weights exploited in Cupid, in XClust and in the approach of \cite{YiHuCh05}. A more complex option is
adopted by {\tt LSD}, where weights are obtained by solving a linear regression problem.

Other options focus on operators like {\tt Max} and {\tt Min}; {\tt Max} returns the highest similarity score between $u$ and $v$ among all available scores; {\tt Min} returns the lowest one.

In \cite{WoMlDo10}, a {\em non-linear combination scheme} is considered. Here, $\lambda$ is defined as follows:

\begin{equation}
\lambda(u,v) = \alpha \cdot \max\{SemSim(u,v),SynSim(u,v)\} +  (1 - \alpha) \cdot CardSim(u,v)
\label{eqn:nonlinearcombination}
\end{equation}

\noindent being $\alpha$ a real number. The functions $SemSim(u,v)$, $SynSim(u,v)$ and $CardSim(u,v)$ have been
introduced in Section \ref{subsub:Element-Level-Matching}.

Further examples of non-linear combination schemas have been provided in \cite{AlNaSa10} and \cite{Jeong*08}. Such
schemas rely on the idea that the results generated by several similarity functions are somewhat related to each other
\cite{AlNaSa10} and, sometimes, partial similarity scores appear to be collinear \cite{Jeong*08}. Therefore, in
general, given two similarity scores $\sigma_1(u,v)$ and $\sigma_2(u,v)$, one can expect that if $\sigma_1(u,v)$ is
high then $\sigma_2(u,v)$ will be high too.

The approach of \cite{AlNaSa10} proposes a definition of $\lambda$ which takes the interdependencies among partial
similarity scores into account. Such a definition, according to our notation, is as follows:

\begin{equation}
\label{eqn:nonlinear}
{\small
\begin{split}
\lambda(u,v) = \rho \sum_{l =1}^{k}\sigma_l(u,v) \pm (1 -\rho) \sum_{i =1}^{k}\sum_{j =1}^{k} \sigma_i(u,v)\sigma_j(u,v)
\end{split}
}
\end{equation}

In Equation \ref{eqn:nonlinear}, $\lambda(u,v)$ consists of two terms: The former is equal to the sum of the partial
similarity scores and, therefore, it coincides with Equation \ref{eqn:weightedsum} when all weights are set equal to 1,
i.e. when all partial similarity scores equally contribute to the global one. The latter term takes the
interdependencies between partial similarity scores into account. If $\sum_{l =1}^{k}\sigma_l(u,v)$ is greater than a
threshold, the first and second terms are added (i.e., the symbol $+$ is used in Equation \ref{eqn:nonlinear}). This
means that the interdependencies between pairs of partial similarity scores are used to reinforce the global one. By
contrast, if $\sum_{l =1}^{k}\sigma_l(u,v)$ is lower than a threshold, the second term is subtracted to the first one.
As usual, the coefficient $\rho$ is instrumental in normalizing $\lambda(\cdot,\cdot)$ to the real interval $[0,1]$.

\subsection{The schema similarity function $\phi$}
\label{sub:phi}

In this section we investigate the main features of the $\phi$ function. As emerges from Definition
\ref{def:xmlmatcher}, the role of $\phi$ is to compute the similarity degree of a pair of DTDs/XSDs. As we will show
later, this is a key step to perform Schema Clustering. However, we point out that the definition of $\phi$ is
optional, and, therefore, some of the approaches discussed till now do not provide any definition for $\phi$.

In \cite{AlScSa08}, the authors consider a group of XSDs $\mathcal{S} = \{S_1, S_2, ..., S_n\}$, and, for each pair
$\langle S_i, S_j\rangle$ of schemas, they apply a structural and a linguistic Matcher. Therefore, each pair $\langle
u,v \rangle$ of schema elements, such that $u \in S_i$ and $v \in S_j$ has associated a linguistic similarity score
$sc_l(u,v)$ and a structural similarity score $sc_s(u,v)$. The overall similarity score of $\langle u, v \rangle$ is
obtained by computing the weighted mean of $sc_l(u,v)$ and $sc_s(u,v)$. The overall similarity score of $\langle S_i,
S_j \rangle$ is obtained by summing up all the overall similarity scores of the pairs $\langle u, v \rangle$ such that
$u \in S_i$ and $v \in S_j$.

\begin{equation}
\label{eqn:algergawy}
{\small
\phi(S_i,S_j) = \sum_{u \in S_i} \sum_{v \in S_j} \alpha sc_l(u,v) + \left(1 - \alpha\right) sc_s(u,v)
}
\end{equation}

In XIKE, semantic matchings are used to map XSDs onto points of a multidimensional space \cite{Ursino-JODS1}. In
detail, given a set $\mathcal{S} = \{ S_1 ,S_2, \ldots, S_n \}$ of XSDs, a {\em multi-dimensional vector space} $V$ is
built so that each element appearing at least in one of the XSDs of $\mathcal{S}$ corresponds to a dimension. Two
elements involved in a semantic matching are collapsed into a unique dimension in $V$. An XSD $S_i \in \mathcal{S}$ can
be described on the basis of its elements. As a consequence, it can be mapped onto an array $\vec{v}_i \in \R^N$, being
$N$ the overall number of the complex elements of all the XSDs of $\mathcal{S}$. The computation of the similarity
between two XSDs $S_i$ and $S_j$ reduces to the computation of the {\em Euclidean distance} between the corresponding
vectors $\vec{v}_i$ and $\vec{v}_j$.

In XClust, the computation of the similarity degree of two DTDs is performed by computing the similarity degree of the
trees associated with them. In detail, given a collection $\mathcal{D} = \{D_1,\ldots,D_n \}$ of DTDs, XClust considers
all pairs of DTDs in $\mathcal{D}$ and, for each pair, it finds the pairs of {\em similar elements} between the two
DTDs. More formally, given two DTDs, $D_i \in \mathcal{D}$ and $D_j \in \mathcal{D}$, and a threshold $\tau$, a list of
{\em similar elements} $L_{ij}$ is built; such a list consists of all the pairs $\langle u, v \rangle$ of elements such
that $u \in D_i$, $v \in D_j$ and the similarity degree of $u$ and $v$ is greater than $\tau$. The role of $\tau$ is,
therefore, to filter out those pairs of elements whose similarity degree is not recognized as sufficiently high. The
similarity degree of $D_i$ and $D_j$ is proportional to the size of $L_{ij}$. However, it can happen that a DTD is
similar to a subpart of a larger DTD (think, for instance, of a DTD describing a University Department and a DTD
describing the whole University). In order to correctly handle this scenario, $|L_{ij}|$ must be normalized; for this
purpose, it is divided by a suitable coefficient that, in case of XClust, is $\min\left(|D_i|,|D_j|\right)$.

In \cite{WoMlDo10}, the similarity degree of two DTDs is determined by computing the edit distance \cite{Tai79} between
two simplified trees. In detail, as observed in Section \ref{sub:mapping}, the approach of \cite{WoMlDo10} maps each
DTD $D_i$ onto a tree $T_i$. Since such a mapping is a bijection, in the following we will use the symbols $D_i$ and
$T_i$ in an interchangeable fashion and, therefore, when we will mention a tree $T_i$ we will implicitly refer to the
DTD $D_i$ generating it. Some primitive operations, called {\em edit operations}, on $T_i$ are allowed \cite{NiJa02};
among them we cite the {\em re-labelling} of a node (i.e., the label of a node $v$ in $T_i$ is changed from $l_v$ to
$l^{\prime}_v$), the insertion/deletion of a node at a specified level and the insertion/deletion of a subtree in
$T_i$. In the following each of these edit operations will be denoted in short as $\mathtt{op}$ and we assume that a
function (called {\em cost function}) is defined. Such a function associates each allowed operation $\mathtt{op}$ with
a non-negative real number (called {\em cost}). Let us consider a pair of trees $T_i$ (called {\em source}) and $T_j$
(called {\em target}). An {\em edit script} $\mathtt{es}$ is defined as a sequence $\mathtt{es} = \{\mathtt{op}_1,
\ldots, \mathtt{op}_k \}$ of edit operations allowing $T_i$ to be transformed into $T_j$. The cost
$\gamma(\mathtt{es})$ of an edit script $\mathtt{es}$ is defined as the sum of the costs of the operations forming
$\mathtt{es}$: So, for instance, if an edit script would only consist of the insertion of a node $v$ in $T_i$ and the
re-labelling of another node $w$, then its cost would be equal to the sum of the costs of these two operations. In
principle, there could be several edit scripts from $T_i$ to $T_j$, and each of them has its own cost. An {\em optimal
edit script} from $T_i$ to $T_j$ is an edit script from $T_i$ to $T_j$ having the minimum cost; its cost represents the
the tree-edit distance from $T_i$ to $T_j$. Once again, observe that several optimal edit scripts may exist from $T_i$
to $T_j$; however, even in this case, the tree edit distance is uniquely defined. According to the definition above,
the equation for defining the tree edit distance (and, the function $\phi(S_i,S_j)$ which coincides with it) is as
follows:

\begin{equation}
\label{eqn:phieditdistance}
\begin{split}
\phi(D_i,D_j) = & \min\{\gamma(\mathtt{s}) | \mathtt{s} \text{ is an edit script from $T_i$ to $T_j$} \}
\end{split}
\end{equation}

In Table \ref{tab:phicomparison} we summarize the main features of $\phi$ for the approaches discussed in this section.

\begin{table*}[th!]
\centering

{\scriptsize
\begin{tabular}{||c|c|c|c|c|c||}
\hline  \hline
{\em Approach}                      & {\em Input} & {\em Mapping} & Definition of $\phi$ & {\em Threshold} & {\em Time}       \\
                              & {\em Data}  &         &        &           & {\em Complexity}  \\

\hline \hline
Approach of \cite{AlScSa08}   & XSDs     &   -     & $\phi(S_i,S_j) = \sum_u\sum_v \alpha sc_l(u,v) +$ & $\alpha \in [0,1]$ & $O(n_i \cdot n_j)$ \\
                              &         &         & $+ \left(1 - \alpha\right) sc_s(u,v)$             &                     & $n_i = |\mathcal{E}(S_i)|, n_j = |\mathcal{E}(S_j)| $          \\

\hline
XIKE                          & XSDs     &  Each XSD  $S_i \in \mathcal{S}$       & $\phi(S_i,S_j) = \sqrt{\sum_{i = 1}^N \left(\vec{v}_i - \vec{v}_j\right)^2}$& - &$O(N)$  \\
                              &  &  is mapped onto a vector     &   &                      & $N = \sum_{S_i \in {\cal S}} |\mathcal{E}(S_i)|$  \\
                              &         &  $\vec{v}_i \in \R^N$ &   &    &                     \\

\hline

XClust                          & DTDs     &  Each DTD $D_i \in \mathcal{D}$  & $\phi(D_i,D_j) = \frac{L_{ij}}{\min\left(|D_i|,|D_j|\right)}$    & $\tau \in [0,1]$  & $O(n_i \cdot n_j)$    \\
                                &     &  is mapped onto a tree $T_i$        &  &                      & $ n_i = |\mathcal{E}(S_i)|, n_j = |\mathcal{E}(S_j)|$\\
                                &         &    &   &  &    \\

\hline

Approach of \cite{WoMlDo10}     & DTDs     &  Each DTD $D_i \in \mathcal{D}$       & See Equation \ref{eqn:phieditdistance}   & - & $O(n_i \cdot n_j)$  \\
                                &      &  is mapped onto a tree $T_i$        &   &   & $ n_i = |\mathcal{E}(S_i)|, n_j = |\mathcal{E}(S_j)|$ \\
                                &         &   &             &   &                        \\

\hline
\hline

\end{tabular}
}

\caption{A comparison of the strategies for the computation of the function $\phi$ for the XML Matchers into
evaluation.} \label{tab:phicomparison}

\end{table*}

\section{Commercial XML Matchers}
\label{sec:commercial}

In this section we review some commercial tools designed to handle Schema Matching tasks; in particular, we discuss how
they handle matchings in DTDs/XSDs. The systems we consider are IBM Infosphere Data Architect, Microsoft Biztalk
server, SAP NetWeaver Process Integration, Harmony, JitterBit and Altova MapForce.

We used the XML Matcher template to compare these approaches and, for each system, we specify its $\mathtt{IFD}$ and provide some discussion on $\mathtt{IRD}$, the pre-processing function $\mathcal{M}$, the similarity function $\sigma$ and the aggregating function $\lambda$.
To the best of our knowledge, none of these systems implements the $\phi$ function.


\begin{table*}[t]
\centering

{\scriptsize
\begin{tabular}{||c|c|c|c|c|c|c|c||}
\hline  \hline
{\em System}         & {\em Reference}   & {\em Vendor}  & $\mathtt{IFD}$ &{\em Element vs.} & {\em Match}        &   $\sigma$         & $\lambda$      \\
                     &                   &               &              & {\em Structural}  & {\em Cardinality}  &                      &               \\

\hline \hline
Infosphere     &      -               & IBM & Relational/   & Structural & 1:1 &  Name               &  -  \\
               &                     &     & Metadata      &            &     &  Matching+          &     \\
               &                     &     &               &            &     &  External           &     \\
               &                     &     &               &            &     &  Thesauri           &     \\
\hline
BizTalk        & \cite{BeMeCh06}     & Microsoft & XML     & Element   & $1:1$ & Name               &  Yes  \\
Server         &                     &           &         &           &       & Matching  +        &        \\
               &                     &           &         &           &       & Past User          &        \\
               &                     &           &         &           &       & Matchings          &        \\

\hline
NetWeaver      & \cite{PeEbRa11}     & SAP       & XML/SQL/& Element +   & $m:n$ & Name             & Yes  \\
               &                     &           & OWL     & Structural  &       & Matching +  &      \\
               &                     &     &               &            &     &  External           &     \\
               &                     &     &               &            &     &  Thesauri           &     \\
               &                     &           &         &             &       & Data Type +  &      \\

               &                     &           &         &             &       & Graph-Based      &      \\

\hline
Harmony      &  \cite{Smith*09}    & Open & XML/SQL  & Element +  & $1:1$ &  Tokenization/   & Yes  \\
             &                     & Source& OWL      & Structural &      &  Stemming        &      \\
\hline
JitterBit     &       -               & Open   & XML/       & Element +  & $1:1$ &   Name       & -   \\
              &                      & Source & Relational & Structural &       &   Matching &    \\
\hline
Altova        &       -               &  Altova    & XML & Element +  &$1:1$   &    Name            & -  \\
MapForce      &                      &            &     & Structural &        &    Matching         &   \\
\hline \hline
\end{tabular}
}

\caption{Comparison of commercial XML Matchers in the light of our XML Matcher Template} \label{tab:main-features-commercial}
\end{table*}

In some cases, the design and the development of these systems were performed in a strict integration with academic
researchers. This is, for instance, the case of NetWeaver, which is strongly based on the COMA++ prototype
\cite{Aumuller*05}, and of Auto Mapping Core ({\em AMC}) \cite{PeEbRa11}, a framework conceived for supporting the
integration of multiple schema matching approaches, which is a nice example of research collaboration between SAP and
the University of Leipzig.

For each system we indicate the company commercializing it, the possible references to academic papers associated with
it, and if it is either distributed as an open source suite or not. It is worth observing that Harmony was developed by
MITRE corporation but it was used in conjunction with the AquaLogic Data Service platform \cite{Carey*07} (initially
developed by BEA systems and subsequently acquired by Oracle Corporation in 2008).

As for the {\em input format domain} (which corresponds to the column labeled as $\mathtt{IFD}$), we observe that, in
most cases, these systems are able to manage a wide range of data sources (like OWL Schemas, SQL DDL schemas, EDI and
flat files), but all of them are able also to deal with DTDs/XSDs.

We observe that most of the technical documents available for these systems do not provide any explicit information
about $\mathtt{IRD}$ and, then, on the internal format used for representing schemas to match.

The next 3 columns of Table \ref{tab:main-features-commercial} focus on the $\sigma$ function and, in particular,
specify if the matching strategy operates at the element level or at the structural one, report the cardinality of
discovered matchings and, finally, the algorithm used to discover matchings.

As for the underlying matching algorithms, almost all of them use information at both the element level and the
structural one.

As for the cardinality of the matchings that can be discovered by analyzed systems, most of them support only $1:1$
matchings, even if BizTalk Server can discover also $1:n$ matchings and SAP NetWeawer handles also $m:n$ matchings.

All the commercial systems discussed in this section provide end-users with GUIs allowing them to specify matchings
between schemas (see below for a detailed discussion). In addition, they offer (often limited) capabilities to find
matchings in a semi-automatic fashion. According to Table \ref{tab:main-features-commercial}, the simplest matching
technique requires to check that two schema elements share the same name. Name Matching is implemented by all the
systems, even if each of them supports also other and more sophisticated techniques: For instance, InfoSphere uses
external thesauri, NetWeaver takes advantage of constraints about element cardinalities, and Harmony applies string
pre-processing procedures, like tokenization/stemming. It is also interesting to observe that BizTalk Server provides a
function to rank pairs of candidate matchings. This function is based on two heuristics: The former leverages on
lexical similarities and element types, whereas the latter considers past user matching actions.

Some systems, e.g. NetWeaver and Harmony, implement an aggregation function $\lambda$, i.e. they exploit multiple
Matchers to generate semantic matchings and properly aggregate the results generated by each matcher. The function
$\lambda$ implemented in these systems is, generally, quite simple: for instance, some aggregation operators are the
weighted sum operator (described in Equation \ref{eqn:weightedsum}) or the {\tt MAX} operator.

All the commercial systems discussed in this section require end-users (in general, expert people) to manually specify
semantic matchings between pairs of schema elements. Since involved schemas may be very large, advanced GUIs are
provided to make this process easier \cite{BeMeCh06,PeEbRa11,Smith*09}. In the simplest cases, a GUI reports the
schemas to match and uses a ``line-drawing'' visualization approach, i.e. in order to specify a matching, the user is
asked to draw a line joining the pair of involved elements.

The GUI provided by BizTalk Server is similar to that offered by Altova MapForce and allows for exporting mappings
specified by users into the XSLT format. The 2010 edition of this system has been significantly improved so as to
better handle large schemas, and supports an enhanced user interface to better visualize complex mappings.

NetWeaver provides the {\em Matching Process Designer}, that not only reports candidate matchings but also helps users
to select the techniques to adopt for finding correct matchings. A further, relevant feature of NetWeawer (and, in
particular, of the Auto Mapping Core component mentioned above) is the opportunity of analyzing the intermediate
results of a matching process (i.e., the user can check whether the output of a matching algorithm fits her desiderata
or not, in which case she can turn to another Matcher).

In Harmony, the GUI supports a variety of filters that help users/data architects in selecting appropriate matchings.

JitterBit focuses on data integration in the context of point-to-point application integration, ETL and SOA. It
consists of two main components, namely an Integration Environment and an Integration Server. The former has a user
friendly GUI for supporting users in the integration process.


\section{Challenges on XML Matchers}
\label{sec:Challenges}

Despite huge efforts in the area of XML-specific Schema Matching have been put both at the academic and the industrial
levels, there are several issues that deserve of being addressed properly.

In this section we focus on two emerging challenging areas, strictly related to the theme of DTDs/XSDs Matching. The
former (illustrated in Section \ref{sub:schemaclustering}) deals with the clustering of DTDs/XSDs. The latter
(discussed in Section \ref{sub:uncertainty}) is about {\em uncertainty} in the matching process.

\subsection{Schema clustering and data integration at Web scale}
\label{sub:schemaclustering}

Approaches to clustering DTDs/XSDs have proven to play a key role in data integration tasks on a Web scale
\cite{Madhavan*07}. In fact, traditional data integration techniques generally assume that all the sources to integrate
belong to the same domain. Unfortunately, in a Web scenario, it is likely to find data sources spanning multiple (and,
in many cases, only partially related) domains.

For instance, consider a scenario in which we want to realize a unique access point for all the services provided
through portals handled by the Central Government Offices of a country. The categories of the involved portals and the
number of portals for each category are reported in Table \ref{tab:schemaclustering}.

\begin{table}
\centering

{\scriptsize
\begin{tabular}{||c|c||}
\hline \hline
{\em Portal categories} & {\em Number of portals}  \\
\hline \hline
Statistics & 1 \\
\hline
Representations & 2 \\
\hline
Peripherical Offices & 6 \\
\hline
Certifications & 4 \\
\hline
Property Register Offices & 3 \\
\hline
Social Security & 9 \\
\hline
Foreign Relations & 4 \\
\hline
Relations Abroad & 6 \\
\hline
Defense & 1 \\
\hline
Justice & 10 \\
\hline
Criminality & 6 \\
\hline
Internal Security & 6 \\
\hline
Public Assistance & 5 \\
\hline
Health Services & 6 \\
\hline
Education & 3 \\
\hline
Environment & 8 \\
\hline
Cultural Goods & 10 \\
\hline
Employment & 9 \\
\hline
Farms & 3 \\
\hline
Industrial Companies & 9 \\
\hline
Transportations & 10 \\
\hline \hline
\end{tabular}
}
\caption{Categories and numbers of involved portals}
\label{tab:schemaclustering}
\end{table}

Clearly, if we integrate the DTDs/XSDs of all these portals in one time to construct the virtual schema of the access
point, the overall global virtual schema would be enormous, heterogeneous, confused, and, ultimately, unusable for
citizens.

\begin{figure}
\centering
\includegraphics[width=11cm]{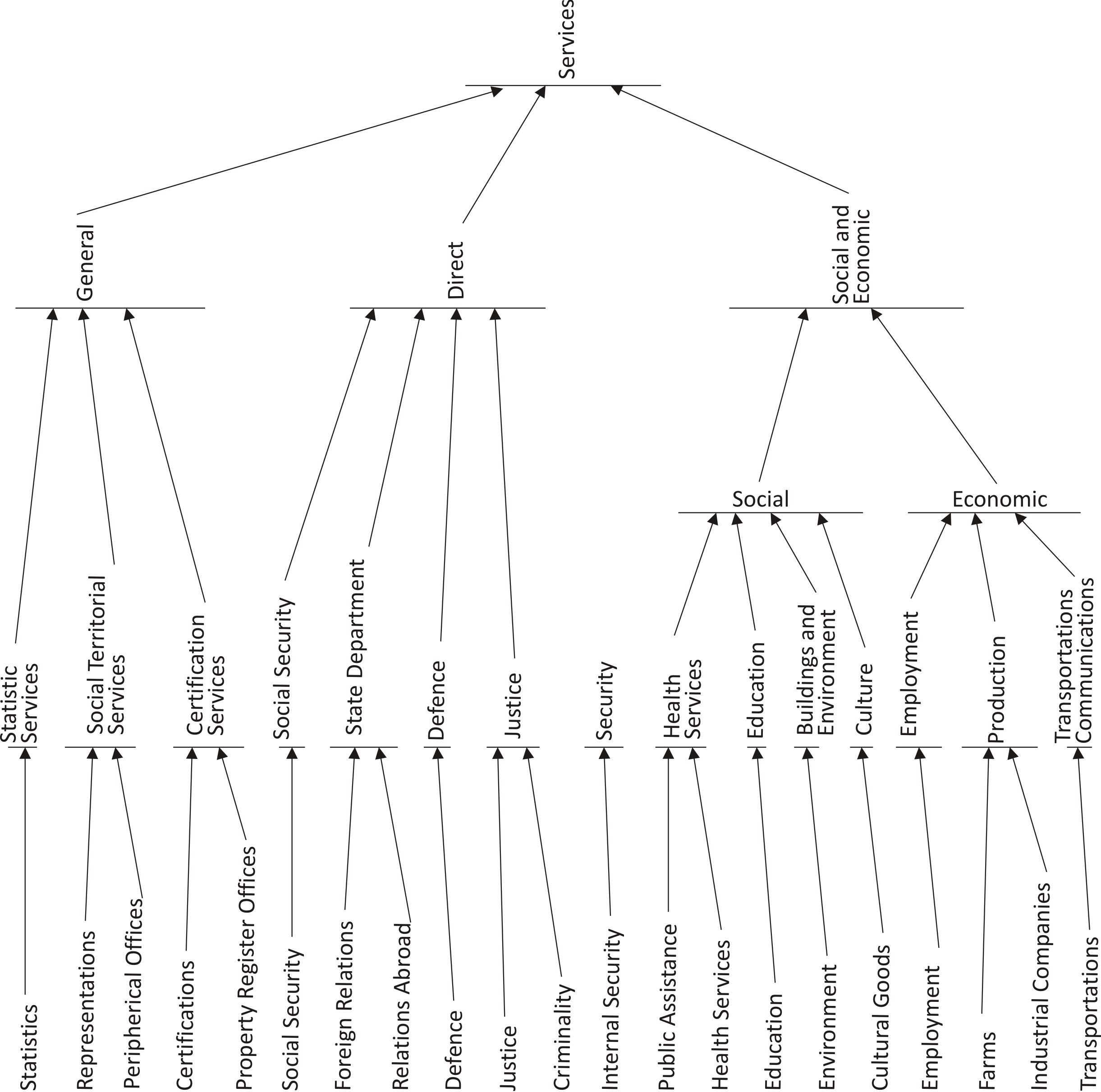}
\caption{An example of the role of clustering in data integration at Web scale}
\label{fig:schemaclustering}
\end{figure}

A better solution to this problem would require that first available DTDs/XSDs are clustered and classified into
homogeneous domains, then the DTDs/XSDs of each domain are integrated in such a way as to obtain a DTD/XSD representing
the domain in the whole; after this, the DTDs/XSDs of detected domains are clustered in such a way as to obtain more
abstract domains, each represented by a unique DTD/XSD. This process could be repeated until to a unique very abstract
DTD/XSD, representing all available services, is obtained. This would be the schema associated with the access point.
Figure \ref{fig:schemaclustering} provides a representation of this approach.

Interestingly, this way of proceeding can be considered as a generalization of the schema clustering techniques for
automatically classifying existing DTDs/XSDs (and, in general, other kinds of data source) into classes
\cite{HeTaCh04,MaAb10}. It has a threefold effect: {\em (i)} the {\em data integration task is more effective}, because
only conceptually related sources are involved in it; {\em (ii)} {\em the data integration task is faster}, because the
number of sources in a cluster is significantly smaller than the overall number of sources; {\em (iii)} {\em the data
querying activity is quicker, more efficient and effective} because the user is guided to find information of her
interest.

Schema clustering approaches can be classified into three categories, i.e. {\em Mapping-Based}, {\em Tree-Based} and
{\em Multi-Strategy} approaches.

Mapping-Based approaches consider DTDs/XSDs as {\em points of a high dimensional space}, map them onto {\em points of a
lower dimensional space} and define the similarity of two DTDs/XSDs as the distance of the points representing them in
the lower dimensional space. This information is, then, used by a clustering algorithm. Examples of approaches
belonging to this category are XIKE and the approach of \cite{Qian*00}. XIKE operates on XSDs; it has been discussed in
Section \ref{sub:phi}. The approach of \cite{Qian*00} works on DTDs. In a first stage, it clusters the elements of all
available DTDs on the basis of their linguistic similarity. Such a clustering task can be interpreted as a {\em
dimensionality reduction} activity in which the set of the elements generating the DTDs is mapped onto a lower
dimensional space. As in XIKE, the mapping task is performed with the support of discovered semantic matchings. In this
new space, an arbitrary DTD can be represented as an array having one component for each cluster; the $i^{th}$
component of this array indicates how many elements of the corresponding DTD belong to the $i^{th}$ cluster.
Experimental tests showed that these approaches are able to achieve a high accuracy level; however, the preliminary
mapping phase plays a key role, and possible inaccuracies in it can negatively influence the whole clustering process.

Tree-Based approaches represent DTDs/XSDs into examination as trees and use linguistic and context-level Matchers to
find their similarities. As observed in Section \ref{sub:phi}, XClust and the approach of \cite{WoMlDo10} use the
tree-based representation of DTDs to compute their similarity degree. A further example of these approaches is {\em
XMine} \cite{NaIr07}. {\em XMine} exploits WordNet, in conjunction with a user-defined dictionary, to find semantic
matchings. Once schema similarities have been computed, an arbitrary clustering algorithm can be applied. Tree-Based
approaches are effective because they consider both the structure and the content of a DTD/XSD. However, schemas may be
large and the computation of the similarity degree of two schemas may require the corresponding trees to be traversed
more times. Such an operation can be computationally expensive. For this reason, some authors \cite{AlScSa08} propose
to reduce computational costs by substituting trees with Prufer sequences \cite{Prufer18}.

{\em Multi-Strategy approaches} use multiple algorithms to compute the matching degree between any element of a source
schema and any element of a target one. Then, they combine obtained results to generate a global similarity score for
each pair of involved schemas. A relevant example of Multi-Strategy approaches is {\em Schemr} \cite{ChMaHa09}. Schemr
is a {\em Schema search engine}, i.e. it offers functionalities for searching and retrieving schemas from a given
repository. Retrieved schemas could be relational ones or XSDs. In Schemr, users (typically data\-base administrators)
can submit a query (consisting of a list of keywords) representing their needs. Furthermore, they can specify a schema
fragment and require Schemr to find, in the repository, other schemas which resemble (and, potentially, complete) it.
For this purpose, Schemr first retrieves a set of {\em candidate schemas}, which are likely to fit user query, and then
applies an {\em ensemble of Matchers} on them. Each Matcher produces a {\em similarity matrix} that reports the
similarity degree between any element of the user query and any element of a given candidate schema. The similarity
matrices produced by each Matcher are finally combined to generate a {\em global similarity matrix}, and candidate
schemas are ranked according to their similarity with respect to user query. A further example of Multi-Strategy
approach is {\em Affinity} \cite{Smith*10}. It first extracts a set of keywords from each involved schema, after having
applied some filters, like stopwords removal or stemming algorithms, on them. Extracted keywords are used to summarize
the content of a schema. Afterwards, it constructs the distance between each pair of schemas by computing the Jaccard
coefficient of the corresponding set of keywords. Finally, it uses this information as an input to a hierarchical
clustering algorithm.

\subsection{Uncertainty management in XML Matchers}
\label{sub:uncertainty}

In the latest years, an emerging and widespread belief in the Schema Matching research area is that the output
generated by a Matcher is often {\em uncertain}. To describe this problem we introduce the following example.

\begin{figure}[h]
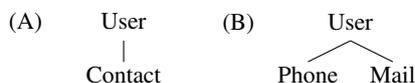
%
	\centering
	\small
	\begin{minipage}{3cm}
		\begin{center}
        \small (A)
        \smallskip
		\Tree [.{User} [.{Contact} ]]		
		\end{center}		
	\end{minipage}
	\begin{minipage}{3cm}
		\begin{center}
		\small (B)
        \smallskip
        \Tree [.{User} [.{Phone} ] [.{Mail} ]]		
		\end{center}		
	\end{minipage}
	\caption{Two fragments of XSDs representing the contacts of a user.}
	\label{fig:xmluncertainty}%
\end{figure}

\begin{example}
\label{ex:uncertainty}
{\em Let us consider two fragments of two XSDs describing the contact details of a Student who is entitled of borrowing
books from a University Library. The first fragment is reported in Figure \ref{fig:xmluncertainty}(A); here the {\tt
User} element has the sub-element {\tt Contact}. The second fragment is reported in Figure \ref{fig:xmluncertainty}(B);
in this case, the {\tt User} element has the two sub-elements {\tt Mail} and {\tt Phone}.
}
$\punto$
\end{example}

Assume to run a Matcher in order to find semantic matchings between the first and the second fragment. The Matcher
could associate the same similarity score with the pairs $\langle {\tt Contact}, {\tt Phone} \rangle$ and $\langle {\tt
Contact}, {\tt Mail} \rangle$. Such a configuration is intrinsically uncertain because, in principle, we do not know if
the matching $\langle {\tt Contact},$ ${\tt Phone} \rangle$ is preferable to the matching $\langle {\tt Contact}, {\tt
Mail} \rangle$.

A possible solution consists of requiring human advice to identify actual matchings.

In the absence of it, a potential solution would consist of listing all potential matchings and selecting the one
having the highest score. Unfortunately, such a solution can yield to information loss. For instance, assume we want to
know the contact details of all the students who borrowed a given book. Depending on the selected matching, different
results will be obtained: If we assume as true the matching involving {\tt Contact} and {\tt Mail} we loose those users
whose {\em only information} contact is given by phone number. By contrast, if we would opt for the matching involving
{\tt Contact} and {\tt Phone}, we would fail to retrieve those users whose information contact is given only by mail
address.

In order to manage uncertainty in Schema Matching process, several authors introduced the concept of {\em probabilistic
mapping}, i.e. they suggested to associate a score with each discovered matching, stating its probability of being
correct \cite{Gal11,DaDoHa08,DoHaYu09}.

In \cite{DoHaYu09} the authors studied uncertain matchings, but in the context of relational databases. They analyzed
the complexity of evaluating Selection-Projection-Join (SPJ) queries in presence of uncertainty. Gal et al.
\cite{Gal*09} extended this approach to aggregation operators (e.g., {\tt COUNT}).

We refer the reader to the excellent book by Gal \cite{Gal11} for an update survey on this topic.

The real problem is that the uncertainty management in the context of XML sources received much less attention and, to
the best of our knowledge, there is only one approach, due to Cheng {\em et al.} \cite{ChGoCh10}, subsequently extended
in \cite{GoChCh12}, that deals with this issue. In detail, given a pair of DTDs/XSDs $S$ ({\em source schema}) and $T$
({\em target schema}), the authors of \cite{GoChCh12} assume that one element $u$ of $S$ can form a matching with {\em
at most} one element $v$ of $T$. Such a matching is also called {\em correspondence}. A set of correspondences between
the source and the target schema is called {\em mapping}. As observed before, for each mapping, it is possible to
compute the probability of its correctness. The number of mappings between $S$ and $T$ is exponential against the
number of nodes in $S$ (because any possible subset of the nodes of $S$ can form a mapping), and this poses
computational and storage challenges.

In fact, on one hand, an efficient procedure is required to extract mappings: For this purpose, the authors of
\cite{GoChCh12} suggest to find only the $h$ mappings with the highest probability, being $h$ an ad-hoc parameter. This
requires to solve an $h$-maximum bipartite matching problem \cite{RoGaDo08,Gal2006}. Unfortunately, this problem
requires the management of large bipartite graphs, and this is still computationally demanding. In order to reduce the
computational cost, the bipartite graph is recursively partitioned in a collection of smaller bipartite graphs, and a
maximum bipartite matching algorithm is applied on each of them. The results of these algorithms are finally merged to
produce the top-$h$ mappings.

On the other hand, the space required for storing all derived mappings can be very large. In order to face this
problem, the authors of \cite{GoChCh12} observe that available mappings generally present a high overlap, in the sense
that different mappings could contain the same correspondence(s). As a consequence, all the correspondences shared by a
sufficiently large number of mappings can be stored in an array called {\em block}. Blocks are managed by means of a
tree (called {\em block tree}). In this tree, each node coincides with a node in the target schema and points to a
block storing all the correspondences involving it. The block tree is also useful to efficiently evaluate queries (in
particular, {\em twig queries}) by including uncertain mappings. For this purpose, a query is recursively decomposed
into multiple sub-queries taking the structure of the block tree into account.

\section{Discussion}
\label{sec:discussion}

The template described in this paper allowed us to shed some light into the commonalities between seemingly unrelated
Matchers as well as to identify differences between approaches that are seemingly similar. In this section we want to
summarize some facts of our analysis that, in our opinion, could be used in the design/development of new XML Matchers.
The main facts we highlight are the following:

\begin{enumerate}

\item {\em Usage of existing matchings}. As observed in \cite{RaBe01}, existing matchings can be reused to improve
    the overall quality of the matching process. In the light of our analysis, the idea of reusing matchings has
    been explored in the context of DTD/XSD matchings, even if further research efforts are required. In
    particular, we point out the existence of standardized/commonly used namespaces which make the design of
    DTDs/XSDs easier. For instance, in Section \ref{sub:reuse}, we observed that the XML Schema paradigm offers
    several methods (e.g., {\em shared components}) allowing the import/reuse of an XSD element defined in other
    XSDs. In our opinion, a promising strategy consists of finding matchings between elements defined in these
    namespaces and, next, to reuse these matchings when it is necessary to find the matchings of two XSDs. To
    clarify this concept, let us focus on the e-business domain and let us consider two popular standards for
    creating XSDs and XML documents, namely XCBL\footnote{{\tt www.xcbl.org}} and OpenTrans\footnote{{\tt
    www.opentrans.org}}. Let $S_1$ and $S_2$ be two XSDs; suppose that $S_1$ contains an element $e_1$ already
    defined in XCBL, whereas $S_2$ contains an element $e_2$ defined in OpenTrans. Assume, now, that some semantic
    matchings are available between XCBL and OpenTrans; in particular, suppose that a matching exists between $e_1$
    and $e_2$; the pair $\langle e_1, e_2 \rangle$ forms a semantic matching between $S_1$ and $S_2$ and it can be
    used to find further matchings.

    A similar strategy has been explored by SAP in the context of the Warp 10 project, with a special emphasis on
    the reuse of semantic matchings between XSDs in the business domain. The key idea is to build a global
    repository $G$ (called {\em consolidated data model}). Such a repository initially consists of standard
    business XSDs; these XSDs are, for instance, those provided by SAP or by other commercial actors. If new XSDs
    are available, they can be integrated in a wiki-like fashion (i.e, data modelers can collaborate to find
    matchings and each data modeler can reuse data types defined by the others). However, in this activity, the
    control of a domain expert is compulsory. If we need to match two XSDs $S_1$ and $S_2$, then we first consider
    the matchings between $S_1$ and $G$ and the matchings between $S_2$ and $G$. Next, these matchings can be
    composed to generate matchings between $S_1$ and $S_2$. However, to the best of our knowledge, the details of
    how SAP performs this composition are not publicly disclosed.

    Due to the proliferation of public XSDs and their increasing use in several domains, like e-business and life
    science, the policy of identifying matchings between these XSDs and of reusing them in the matching of other
    XSDs should be encouraged. Therefore, a new XSD matching algorithm should explicitly consider the reuse of
    semantic matchings between commonly used or standardized namespaces.

\item {\em Integrity Constraints and the Usage of {\tt KEYREFS}}. Most data models explicitly support integrity
    constraints: For instance, a basic example of integrity constraints is given by foreign keys in a relational
    schema; analogously, {\tt ID}/{\tt IDREF} and {\tt KEY}/{\tt KEYREF} pairs are useful to model integrity
    constraints in DTDs/XSDs\footnote{It is worth pointing out that {\tt KEY}/{\tt KEYREFS} are valid only for
    XSD.}. Integrity constraints link a {\em source} element of a schema to a target {\em one}: For instance, a
    single {\tt IDREFS} attribute can reference multiple {\tt ID} elements in a DTD and this is useful to model
    $1:n$ relationships. Some authors \cite{MaBeRa01} suggest to use integrity constraints in the matching process
    but, till now, most of the existing studies focus on relational schemas; for instance, in this context, a form
    of similarity can be recognized between two tables if they are linked by some referential constraints. In the
    context of XSDs, XIKE explicitly considers the usage of {\tt KEYREF}: in particular, in the graph associated
    with an XSD, a {\tt KEYREF} constraint is modeled as an edge connecting two nodes. We argue that referential
    constraints are useful to improve the accuracy of the schema matching approach and, therefore, they should be
    considered in new XML Matchers. The main lesson we learned from our study is that {\tt IDREFS} and {\tt
    KEYREFS} model a structural relationship between schema elements that differs from other kinds of relationship,
    like element/subelement; how to include referential constraints in the matching of two DTDs/XSDs is still an
    open research problem.

\item {\em Computation of schema similarity and the $\phi$ function}. From the analysis of both research prototypes
    and commercial systems, we observed that the computation of similarity between DTDs/XSDs will acquire an
    increasing relevance in the next years: For instance, the computation of schema similarities could be
    propaedeutic to cluster and integrate DTDs/XSDs. However, we recognized that only few of the discussed
    approaches compute schema similarities or, according to the notation introduced in Definition
    \ref{def:xmlmatcher}, just few approaches implement the $\phi$ function. If we look at all the implementations
    of $\phi$ together, it does not emerge any common pattern, and the different implementations of $\phi$ do not
    appear related to each other. As a consequence, we can conclude that none of these approaches is more generic
    (or more specific) than others; for this reason, it is also hard to compare the performance achieved by the
    different approaches. The definition of a unifying theory for the computation of schema similarity is an open
    research problem. The next XML matchers should consider the presence of multiple strategies for implementing
    the $\phi$ function and should provide suitable methods to combine the results produced by each implementation
    of $\phi$.

\end{enumerate}

\section{Conclusions}
\label{sec:conclusions}

Finding semantic matchings between DTDs/XSDs is a key step to ensure a full interoperability across multiple data
sources. In this paper we have provided a detailed analysis of approaches (that we called XML Matchers) explicitly
designed to find matchings between DTDs/XSDs. We discussed in depth the opportunities (which, sometimes, represent also
severe challenges) provided by some constructs which represent key components of the XML specification but are not
available in other data modeling languages, like E/R diagrams or relational schemas (think, for instance, of
distributed namespaces or of the hierarchical organization of schema elements). We introduced a template,
called {\em XML Matcher Template}, to describe the main components of an XML Matcher, their role and their
interactions. We used our template to characterize and compare a set of XML Matchers which gained a large popularity
in the literature. Afterwards, we focused on commercial XML Matchers. Finally, we presented two important challenges
related to XML Matchers, namely the clustering of large collections of DTDs/XSDs and the uncertainty management in XML
Matchers.

Despite Schema Matching and XML Matching are success stories lighted up by brilliant research results, further efforts
are required from both a theoretical perspective and a technological one. On one hand, external sources (like
domain-specific dictionaries or thesauri), which played a significant role in early Schema Matching systems, are now
accompanied by further information sources (like user query logs). Therefore, it becomes compulsory to design and test
new Schema Matchers capable of handling the information offered by these sources in an effective and efficient fashion.
On the other hand, in real-life scenarios, human experts are often part of the matching cycle, because, for instance,
their intervention is required to configure the whole matching workflow or to provide a feedback on the correctness of
matching candidates. Therefore, advanced graphical tools are required to better help them. These tools could support a
human expert to select the fragment of DTDs/XSDs to match, the matching algorithms to apply and the auxiliary data
sources to employ in the matching process. An advancement in these fields would be of great benefit for commercial
systems, which often deal with large DTDs/XSDs containing hundreds of schema elements.

We argue that, in the future, the problems outlined above will catch the interest of researchers in both academy and
industry, and we plan to analyze in detail scientific advancements in these fields.

\section*{Acknowledgments}
We thank the Editor and anonymous Reviewers for their thorough review and highly appreciate the comments and
suggestions, which significantly contributed to improving the quality of our manuscript.

\bibliographystyle{plain}      
\bibliography{bibliografia}   

\end{document}